\documentclass[iop,revtex4,numberedappendix,appendixfloats,twocolappendix]{emulateapj}

\begin{document}
\newcommand{\oi}{\text{[\ion{O}{1}]}}
\newcommand{\oii}{\text{[\ion{O}{2}]}}
\newcommand{\neiii}{\text{[\ion{Ne}{3}]}}
\newcommand{\oiii}{\text{[\ion{O}{3}]}}
\newcommand{\woiii}{\text{$W_\lambda(\oiii)$}}
\newcommand{\nii}{\text{[\ion{N}{2}]}}
\newcommand{\hei}{\text{\ion{He}{1}}}
\newcommand{\heii}{\text{\ion{He}{2}}}
\newcommand{\ha}{\text{H$\alpha$}}
\newcommand{\wha}{\text{$W_\lambda(\ha)$}}
\newcommand{\hb}{\text{H$\beta$}}
\newcommand{\hg}{\text{H$\gamma$}}
\newcommand{\hd}{\text{H$\delta$}}
\newcommand{\he}{\text{H$\epsilon$}}
\newcommand{\hz}{\text{H$\zeta$}}
\newcommand{\hn}{\text{H$\eta$}}
\newcommand{\htheta}{\text{H$\theta$}}
\newcommand{\hiota}{\text{H$\iota$}}
\newcommand{\pa}{\text{Pa$\alpha$}}
\newcommand{\pb}{\text{Pa$\beta$}}
\newcommand{\pg}{\text{Pa$\gamma$}}
\newcommand{\pd}{\text{Pa$\delta$}}
\newcommand{\hi}{\text{\ion{H}{1}}}
\newcommand{\hii}{\text{\ion{H}{2}}}
\newcommand{\hk}{\text{H$\kappa$}}
\newcommand{\caii}{\text{\ion{Ca}{2}}}
\newcommand{\sii}{\text{[\ion{S}{2}]}}
\newcommand{\wlya}{\text{$W_\lambda$({\rm Ly$\alpha$})}}
\newcommand{\wlyaem}{\text{$W_\lambda^{\rm em}$({\rm Ly$\alpha$})}}
\newcommand{\llya}{\text{$L$(Ly$\alpha$)}}
\newcommand{\llyaobs}{\text{$L$(Ly$\alpha$)$_{\rm obs}$}}
\newcommand{\llyaint}{\text{$L$(Ly$\alpha$)$_{\rm int}$}}
\newcommand{\lyafrac}{\text{$f_{\rm esc}^{\rm spec}$(Ly$\alpha$)}}
\newcommand{\lha}{\text{$L$(H$\alpha$)}}
\newcommand{\lhb}{\text{$L$(H$\beta$)}}
\newcommand{\sfrha}{\text{SFR(\ha)}}
\newcommand{\sfrsed}{\text{SFR(SED)}}
\newcommand{\ssfrha}{\text{sSFR(\ha)}}
\newcommand{\ssfrsed}{\text{sSFR(SED)}}
\newcommand{\ebmvneb}{E(B-V)_{\rm neb}}
\newcommand{\ebmvcont}{E(B-V)_{\rm cont}}
\newcommand{\ebmvlos}{E(B-V)_{\rm los}}
\newcommand{\nhi}{N(\text{\ion{H}{1}})}
\newcommand{\lognhi}{\log[\nhi/{\rm cm}^{-2}]}
\newcommand{\lognhitable}{\log\left[\frac{\nhi}{{\rm cm}^{-2}}\right]}
\newcommand{\lya}{\text{Ly$\alpha$}}
\newcommand{\lyb}{\text{Ly$\beta$}}
\newcommand{\lyg}{\text{Ly$\gamma$}}
\newcommand{\comment}[1]{}
\newcommand{\wciii}{\text{$W_\lambda$(\ion{C}{3}])}}
\newcommand{\ciii}{\text{\ion{C}{3}]}}
\newcommand{\interoiii}{\text{\ion{O}{3}]}}
\newcommand{\rsiione}{R(\text{\ion{Si}{2}}\lambda 1260)}
\newcommand{\rsiitwo}{R(\text{\ion{Si}{2}}\lambda 1527)}
\newcommand{\rsii}{R(\text{\ion{S}{2}})}
\newcommand{\siii}{\text{\ion{Si}{2}}}
\newcommand{\cii}{\text{\ion{C}{2}}}
\newcommand{\civ}{\text{\ion{C}{4}}}
\newcommand{\rcii}{R(\text{\ion{C}{2}}\lambda 1334)}
\newcommand{\ralii}{R(\ion{Al}{2}\lambda 1670)}
\newcommand{\qh}{Q(\text{H$^0$})}
\newcommand{\rs}{{\cal R}_{\rm s}}
\newcommand{\fcovhi}{f_{\rm cov}(\hi)}
\newcommand{\fcovmetal}{f_{\rm cov}({\rm metal})}
\newcommand{\fesclya}{f_{\rm esc}^{\rm spec}(\lya)}
\newcommand{\logxi}{\log[\xi_{\rm ion}/{\rm s^{-1}/erg\,s^{-1}\,Hz^{-1}}]}
\newcommand{\logq}{\log[Q/{\rm s^{-1}}]}
\newcommand{\lir}{L_{\rm IR}}
\newcommand{\lbol}{L_{\rm bol}}
\newcommand{\luv}{L({\rm UV})}

\title{A {\em JWST}/NIRSpec Exploration of the Connection between
  Ionization Parameter, Electron Density, and Star-Formation-Rate
  Surface Density in \lowercase{$z=2.7-6.3$} Galaxies}

\author{\sc Naveen A. Reddy\altaffilmark{1},
Michael W. Topping\altaffilmark{2},
Ryan L. Sanders \altaffilmark{3,4},
Alice E. Shapley \altaffilmark{5},
and Gabriel Brammer \altaffilmark{6,7}}

\altaffiltext{1}{Department of Physics and Astronomy, University of California, Riverside, 900 University Avenue, Riverside, CA 92521, USA; naveenr@ucr.edu}
\altaffiltext{2}{Steward Observatory, University of Arizona, 933 North Cherry Avenue, Tucson, AZ 85721, USA}
\altaffiltext{3}{Department of Physics, University of California, Davis, One Shields Ave, Davis, CA 95616, USA}
\altaffiltext{4}{NASA Hubble Fellow}
\altaffiltext{5}{Department of Physics and Astronomy, University of California, Los Angeles, 430 Portola Plaza, Los Angeles, CA 90095, USA}
\altaffiltext{6}{Cosmic Dawn Center (DAWN)}
\altaffiltext{7}{Niels Bohr Institute, University of Copenhagen, Jagtvej 128, 2200 Copenhagen N, Denmark}

\slugcomment{DRAFT:\today}







\begin{abstract}

We conduct a statistical analysis of the key factors responsible for
the variation in the ionization parameter ($U$) of high-redshift
star-forming galaxies based on medium resolution {\em JWST}/NIRSpec
observations obtained by the Cosmic Evolution Early Release Science
(CEERS) survey.  The sample consists of 48 galaxies with spectroscopic
redshifts $z=2.7-6.3$ which are largely representative of typical
star-forming galaxies at these redshifts.  The $\sii$\,$\lambda\lambda
6718, 6733$ doublet is used to estimate electron densities ($n_e$),
and dust-corrected $\ha$ luminosities are used to compute total
ionizing photon rates ($Q$).  Using composite spectra of galaxies in
bins of $\oiii$\,$\lambda\lambda 4960, 5008$/$\oii$\,$\lambda\lambda
3727, 3730$ (i.e., O32) as a proxy for $U$, we determine that galaxies
with higher O32 have $\langle n_e\rangle \simeq 500$\,cm$^{-3}$ that
are at least a factor of $\simeq 5$ larger than that of lower-O32
galaxies.  We do not find a significant difference in $\langle
Q\rangle$ between low- and high-O32 galaxies.  We further examine
these results in the context of radiation- and density-bounded
nebulae, and use photoionization modeling of all available strong
rest-frame optical emission lines to simultaneously constrain $U$ and
oxygen abundance ($Z_{\rm neb}$).  We find a large spread in $\log U$
of $\approx 1.5$\,dex at a fixed $Z_{\rm neb}$.  On the
other hand, the data indicate a highly significant correlation between
$U$ and star-formation-rate surface density ($\Sigma_{\rm SFR}$) which
appears to be redshift invariant at $z\sim 1.6-6.3$, and possibly up
to $z \sim 9.5$.  We consider several avenues through which
metallicity and $\Sigma_{\rm SFR}$ (or gas density) may influence $U$,
including variations in $n_e$ and $Q$ that are tied to metallicity and
gas density, internal dust extinction of ionizing photons, and the
effects of gas density on the volume filling fraction of dense clumps
in $\hii$ regions and the escape fraction of ionizing photons.  Based
on these considerations, we conclude that gas density may play a more
central role than metallicity in modulating $U$ at these
redshifts.

\end{abstract}


\section{Introduction}
\label{sec:intro}

The ionization parameter ($U$)---commonly defined as the ratio of the
number density of incident ionizing photons and the number density of
hydrogen atoms---is a key property of the interstellar medium (ISM)
that depends on the intensity and hardness of the ionizing radiation
field, gas density, and the spatial distribution of gas relative to
ionizing sources.  Accordingly, variations in $U$ can provide
meaningful insights into how the ISM and ionizing sources evolve with
redshift and/or galaxy properties.  Recent observations with the {\em
  James Webb Space Telescope} ({\em JWST}) have revealed the presence
of galaxies with apparently ``extreme'' ionization conditions at the
epoch of reionization, where $U$ may be an order of magnitude or more
larger than that of typical star-forming galaxies at lower redshifts
\citep{tang23, bunker23}.  The strong inverse correlation between $U$
and oxygen abundance ($Z_{\rm neb}$) seen in the local Universe
\citep{dopita86, perez14} is commonly interpreted as a reflection of
the harder ionizing spectra associated with low metallicity stars.
Consequently, the large values of $U$ seen at high redshift are
ostensibly due to the lower metallicity stellar populations
characteristic of galaxies at these redshifts.  While these results
are intriguing, they also motivate a more detailed investigation of
the physical causes of the elevated $U$ inferred for high-redshift
galaxies.

In this context, there have been a number of efforts focused on
explaining the redshift evolution of $U$ at a fixed stellar mass up to
$z\sim 2$.  Early work suggested that this evolution reflects the
decrease in metallicity (and implied increase in hardness of the
ionizing spectrum) of galaxies with redshift at a given stellar mass
(e.g., \citealt{sanders16b, kashino19}).  On the other hand, because
$U$ is sensitive to the total ionizing photon rate (or SFR), the
increase in SFR with redshift at a fixed stellar mass has also been
invoked to explain in part the elevated $U$ seen at high redshift
(e.g., \citealt{kaasinen17}).  Others have pointed to the increased
gas (or SFR) surface densities and/or higher electron densities
($n_e$) as a possible factor in the redshift evolution of $U$
\citep{brinchmann08, liu08, shimakawa15, masters16, bian16, davies21,
  papovich22, reddy23b}.  Previous studies have necessarily been
limited in the redshift range over which one can robustly constrain
$U$ and $n_e$, since the lines typically used to infer them for the
same galaxies are inaccessible from the ground above $z\sim 4$.

The {\em James Webb Space Telescope} ({\em JWST}) now allows access to
the same diagnostics of $U$, $Z_{\rm neb}$, and $n_e$ for galaxies
well into the epoch of reionization, where these properties are
expected to lie outside the ranges typically observed up to $z\sim
2-3$, but which still overlap with the more ``extreme'' sources (i.e.,
with high $U$, low $Z_{\rm neb}$, and high $n_e$) at these lower
redshifts.  Such measurements over larger dynamic ranges in redshift,
stellar characteristics, and physical conditions in the ISM form an
ideal dataset to evaluate the evolution in $U$.  Here, we take
advantage of public {\em JWST}/NIRSpec spectroscopy obtained as part
of the Cosmic Evolution Early Release Science (CEERS) survey to
identify the factors responsible for modulating $U$ in high-redshift
($z=2.7-6.3$) galaxies.

The paper is organized as follows.  The data, line measurements, and
construction of composite spectra are discussed in
Section~\ref{sec:data}.  The empirical relationship between $U$ and
both $n_e$ and $Q$ are presented in Section~\ref{sec:neq}, and these
results are discussed in the context of radiation- and density-bounded
nebulae in Sections~\ref{sec:radiationneb} and \ref{sec:densityneb}.
In Sections~\ref{sec:metallicity} and \ref{sec:sigmasfr}, we consider
several pathways through which metallicity and SFR surface density,
respectively, may affect $U$.  Our conclusions and suggestions for
future work are presented in Section~\ref{sec:conclusions}.  A
\citet{chabrier03} initial mass function (IMF) is considered
throughout the paper.  Wavelengths are reported in the vacuum frame.
We adopt a cosmology with $H_{0}=70$\,km\,s$^{-1}$\,Mpc$^{-1}$,
$\Omega_{\Lambda}=0.7$, and $\Omega_{\rm m}=0.3$.

\section{Data and Measurements}
\label{sec:data}


We used the publicly available NIRSpec Multi Shutter Assembly (MSA)
data from the CEERS program (Program ID: 1345;
\citealt{finkelstein22}; Finkelstein et~al., in prep.; Arrabal Haro
et~al, in prep.) for this analysis.  A total of 318 unique objects in
the AEGIS field were targeted using grating and filter combinations
that yielded medium resolution ($R\sim 1000$) spectroscopy at $\lambda
= 1-5$\,$\mu$m, with an integration of $\simeq 52$\,min per grating
and filter combination.  The NIRSpec data were reduced following the
procedures described in \citet{shapley23a}, \citet{sanders23c}, and
\citet{reddy23a}.  The {\em JWST} \texttt{calwebb\_detector1}
pipeline\footnote{\url{https://jwst-pipeline.readthedocs.io/en/latest/index.html}}
was used to process individual uncalibrated exposures, correcting for
saturated pixels, subtracting bias and dark current, and masking other
artifacts resulting from cosmic ray hits.  Flatfielding, background
subtraction, and wavelength solutions were applied to the 2D
spectrograms.  These spectrograms were then combined using offsets
appropriate for the three-point nod dither pattern.  One-dimensional
spectra were optimally extracted from the spectrograms, as described
in \citet{sanders23c}.  If no emission lines or continua were visible,
then an extraction was not performed.  One-dimensional spectra were
extracted for 252 targets.  Slit loss corrections and a final flux
scaling were applied to force the integrated fluxes from the spectra
to agree with the broadband photometry, details of which can be found
in \citet{reddy23a} and \citet{sanders23c}.

Emission line fluxes were measured by fitting the line profiles with
Gaussian functions, along with a continuum model determined by the
best-fit SED, thus accounting for stellar absorption under the $\hi$
emission lines (see \citealt{sanders23c} for further details).  Most of
the analysis presented here relies on composite spectra that were
generated by shifting each object's spectrum to the rest frame,
converting from flux density to luminosity density, interpolating to a
common wavelength grid, and averaging luminosity densities at each
wavelength point using $3\sigma$ clipping.  Average line luminosities,
line ratios, and their corresponding uncertainties including sample
variance were calculated by generating many realizations of the
composite spectra.  In each realization, the spectra of individual
objects were perturbed according to their measurement errors, and
composite spectra were constructed by random selection of these
perturbed spectra with replacement.  Line flux measurements on the
composite spectra were performed in a manner similar to that of
individual objects.  All line ratios presented here were corrected for
dust attenuation based on the Balmer decrement ($\ha/\hb$) and/or the
higher-order Balmer lines ($\hg$ and $\hd$) for individual objects or
from the composite spectra, assuming the \citet{cardelli89} Milky Way
extinction curve.  This extinction curve has been shown to apply to
typical star-forming galaxies at high redshift \citep{reddy20}.  The
line diagnostics presented in this study are listed in
Table~\ref{tab:ratiodef}, but all available rest-frame optical
emission lines were used in the photoionization modeling discussed in
Section~\ref{sec:metallicity}.

\begin{deluxetable}{lr}
\tabletypesize{\footnotesize}
\tablewidth{0pc}
\tablecaption{Line Diagnostics}
\tablehead{
\colhead{Line Diagnostic\tablenotemark{}} &
\colhead{Definition}}
\startdata
    O3 & $\oiii 5008/\hb$ \\
    O32 & $\oiii 4960+5008/\oii 3727+3730$ \\
    N2 & $\nii 6585/\ha$ \\
    S2 & $\sii 6718+6733/\ha$ \\
    Ne3O2 & $\neiii 3870/\oii 3727+3730$ \\
    R23 & $(\oiii 4960+5008\, + \oii 3727+3730)/\hb$ \\
\enddata
\tablenotetext{}{}
\label{tab:ratiodef}
\end{deluxetable}

As described in \citet{shapley23a}, the broadband photometry of
galaxies in the sample was corrected for the contribution from strong
emission lines and then fit with the stellar population synthesis
models of \citet{conroy09} using the FAST program \citep{kriek09}.
Delayed-$\tau$ star-formation histories were adopted, where ${\rm
  SFR}(t) \propto t \times \exp(-t/\tau)$, with a minimum age of
$10$\,Myr.  Galaxies were fit using two sets of assumptions: the SMC
extinction curve \citep{gordon03} and a subsolar stellar metallicity
($Z_\ast = 0.004$), and the \citet{calzetti00} attenuation curve and a
$\sim$solar metallicity ($Z_\ast = 0.02$) for the redshift and stellar
mass criteria given in \citet{shapley23a}.  For the purposes of this
analysis, we focused on the stellar population ages and stellar masses
derived from the fitting, which are similar to those obtained for the
Binary Population and Stellar Synthesis (BPASS; \citealt{eldridge17,
  stanway18}) models that we consider Section~\ref{sec:xiion}.

\begin{figure}
  \epsscale{1.2}
  \plotone{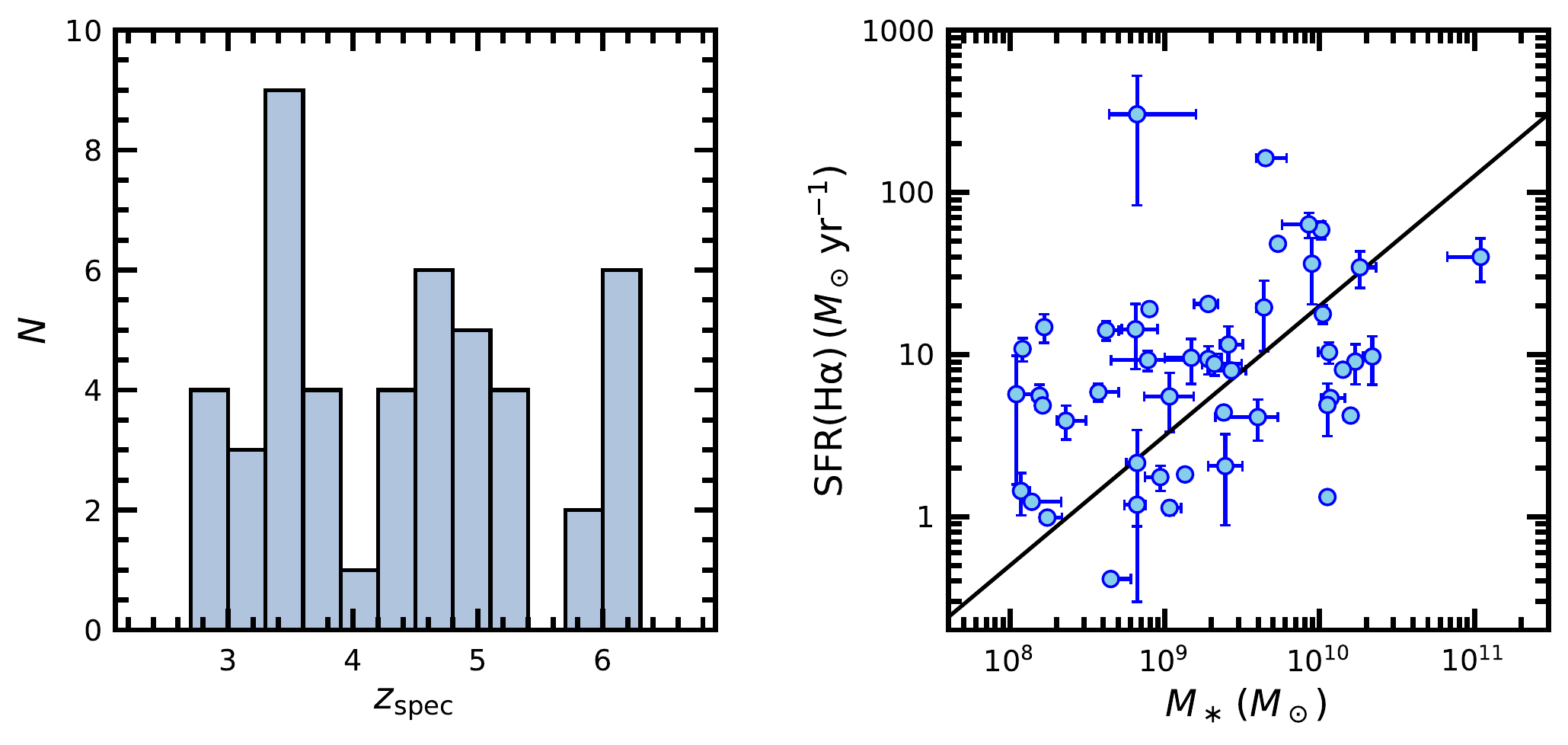}
    \caption{{\em Left:} Spectroscopic redshift histogram of the 48
      galaxies in the sample.  {\rm Right:} Distribution of
      $\ha$-based SFRs and $M_\ast$ for the 48 galaxies in the sample.
      The mean relation between SFR and $M_\ast$ at $z\sim 4.4$
      (roughly the mean redshift of the sample) is indicated by the
      solid line \citep{shapley23a}.}
    \label{fig:sample}
\end{figure}

The final sample used in this study was assembled as follows.  First,
galaxies were required to have secure spectroscopic redshifts $z_{\rm
  spec}= 2.7 - 6.3$, reducing the sample from 252 objects with
extracted spectra to 110.  The lower redshift corresponds to the limit
of ground-based spectroscopic studies of electron densities based on
$\sii$\,$\lambda\lambda 6718, 6733$, the tracer used in this study.
The upper redshift corresponds roughly to the limit where $\sii$ is
accessible with {\em JWST}/NIRSpec.  Second, AGN that were identified
with $\nii/\ha \ge 0.5$ or significant broad components to their
emission lines were excluded, resulting in 106 remaining objects.
Third, we required $\ge 3\sigma$ detections of $\oii$\,$\lambda\lambda
3727, 3730$, $\hb$, $\oiii$\,$\lambda\lambda 4960, 5008$, and $\ha$ to
ensure robust dust corrections and O32, yielding 54 galaxies.
Finally, 6 galaxies which had noisy photometry, thus rendering it
difficult to obtain robust estimates of stellar population properties
(e.g., age, reddening, stellar mass, SFR), were removed from the
sample.  The final sample includes 48 galaxies with the redshift
distribution and SFR and $M_\ast$ shown in Figure~\ref{fig:sample}.
As discussed in \citet{shapley23a}, the bulk of the sample is
representative of star-forming galaxies at $z\la 5$, while above this
redshift the galaxies generally have higher specific SFRs relative to
typical star-forming galaxies (at fixed mass) at those redshifts.

\section{Electron Density and Ionizing Photon Rate}
\label{sec:neq}

For a radiation-bounded (spherical) nebula, the ionization parameter
($U$) is related to the ionizing photon rate ($Q$), hydrogen
density ($n_{\rm H}$), and Str\"{o}mgren radius
($R_{\rm S}$):
\begin{equation}
U \propto \frac{Q}{n_{\rm H}R_{\rm S}^2}.
\label{eq:1}
\end{equation}
The Str\"{o}mgren
radius is also related to $Q$ and the volume-averaged hydrogen density, $\langle n_{\rm H}\rangle$:
\begin{equation}
R_{\rm S}^3 \propto \frac{Q}{\langle n_{\rm H}\rangle^2} \propto \frac{Q}{\epsilon n_{\rm H}^2},
\label{eq:2}
\end{equation}
where the volume filling factor $\epsilon = (\langle n_{\rm
  H}\rangle/n_{\rm H})^2$.  This factor represents the fraction of the
$\hii$ region volume containing denser material that dominates the
emission-line fluxes, such as $\sii$ \citep{osterbrock59,
  kennicutt84}.  Combining Equations~\ref{eq:1} and \ref{eq:2}, and
assuming all the gas is ionized (i.e., $n_{\rm H}\sim n_e$, where
$n_e$ is the electron density), then yields the following dependence
of $U$ on $Q$, $n_e$, and $\epsilon$:
\begin{equation}
U \propto [Qn_e\epsilon^2]^{1/3}.
\label{eq:mainequation}
\end{equation}
Note that $U$ is commonly defined as the ratio of the number density
of ionizing photons over the number density of hydrogen atoms.  For a
radiation-bounded nebula, the number density of ionizing photons is
simply proportional to the photon flux received at the surface of a
sphere with radius $R_{S}$.  An increase in density results in lower
$R_{S}$ and hence a higher photon flux.  Thus, the overall dependence
for a {\em radiation-bounded} nebula is one where $U$ {\em increases}
with $n_{\rm H}$ (or $n_e$).  Motivated by the theoretical
dependencies in Equation~\ref{eq:mainequation}, we first examined how
$U$ varies with both $Q$ and $n_e$.  The photoionization modeling used
to determine $U$ is discussed in Section~\ref{sec:discussion}.  For
the moment, however, we consider the O32 ratio as a proxy for $U$.
The ratio of $\sii$\,$\lambda 6718$ and $\sii$\,$\lambda 6733$,
$\rsii$, was used to infer $n_e$ using the relation from
\citet{sanders16b}.

The ionizing photon rate ($Q$) depends on the SFR and the ionizing
photon rate per unit SFR, or the ionizing production efficiency,
$\xi_{\rm ion}$ \citep{robertson13, bouwens16a, shivaei18, theios19,
  reddy22}, such that $Q \propto \xi_{\rm ion} \times {\rm SFR}$.  The
ionizing production efficiency depends on the properties of massive
stars, such as their stellar metallicity, age, whether they evolve as
single stars or binaries, and the IMF.  For this analysis, $Q$ is
determined empirically from the dust-corrected $\ha$ luminosity
assuming the relationship of \citet{leitherer95} and no escape or dust
absorption of ionizing photons.  We consider the case of a non-zero
escape fraction of ionizing photons from $\hii$ regions in
Section~\ref{sec:densityneb}, and dust absorption of ionizing photons
in Section~\ref{sec:dust}.

\begin{figure}
  \epsscale{1.1}
  \plotone{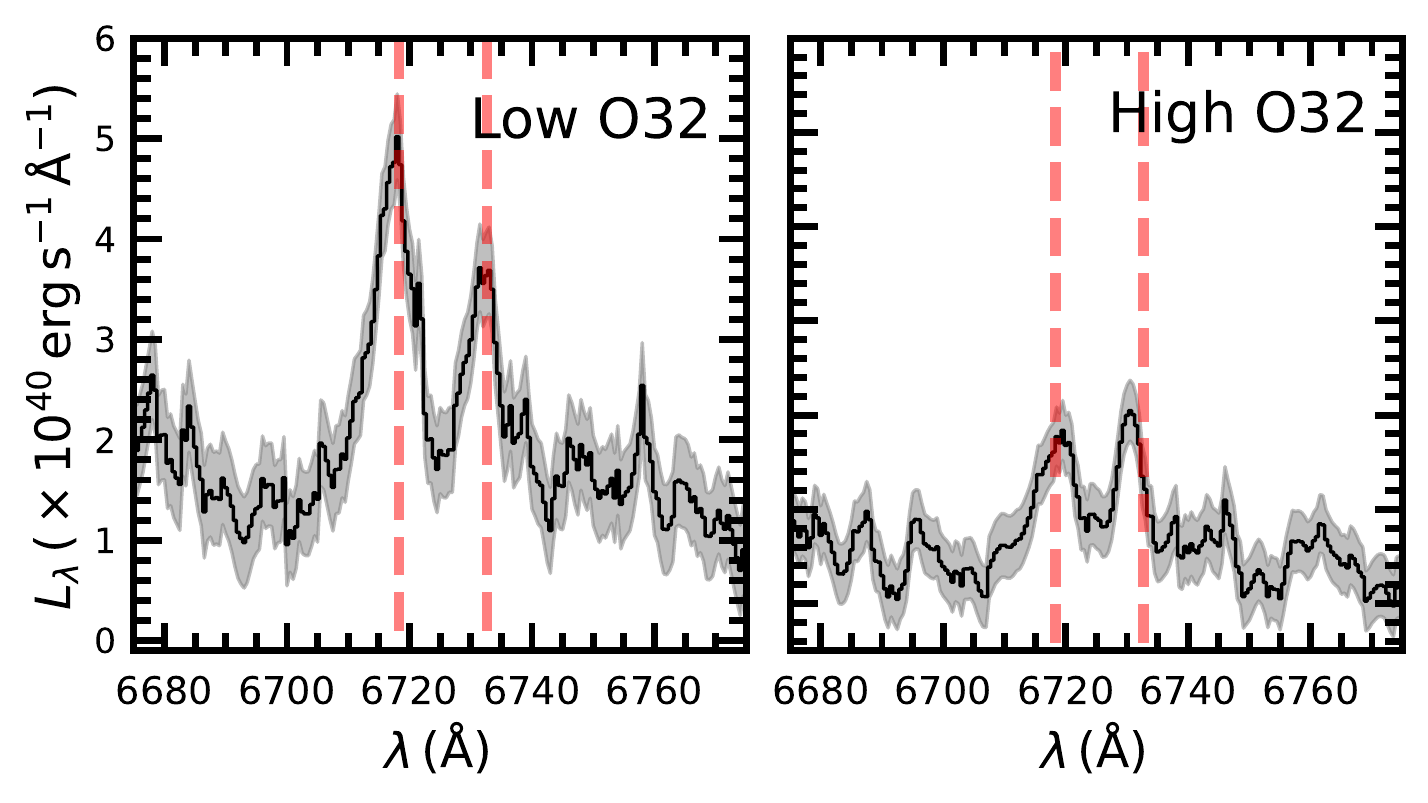}
    \caption{$\sii$\,$\lambda\lambda 6718, 6733$ in the composite
      spectra of the low- and high-O32 bins (left and right,
      respectively), each containing $N=24$ galaxies.  The $1\sigma$
      measurement uncertainty is indicated by the grey shaded
      regions.}
    \label{fig:specplot}
\end{figure}

\begin{figure}
  \epsscale{1.0}
  \plotone{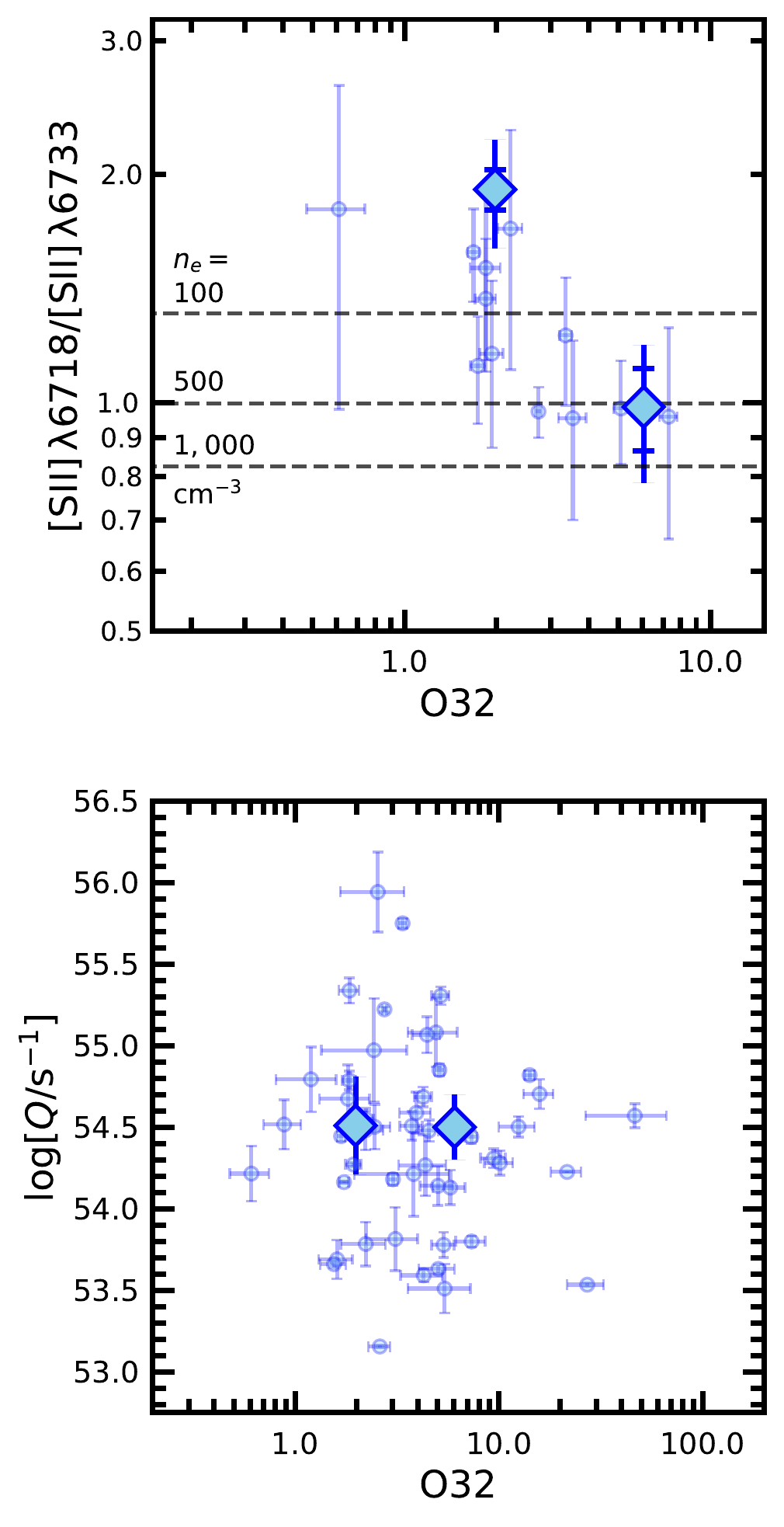}
    \caption{Variation of $\rsii$ (top) and $\logq$ (bottom) with O32.
      Individual galaxies with detections of $\sii$ and $Q$
      measurements are indicated by small circles in the top and
      bottom panels, respectively.  Mean measurements from composite
      spectra of galaxies are shown by the diamonds.  Measurement
      uncertainties for individual and composite values are indicated
      by the capped vertical lines.  Estimates of sample variance
      (including measurement error) for the composite values are
      indicated by the uncapped vertical lines.  In some cases, the
      error bars are smaller than the symbols.}
    \label{fig:densityq}
\end{figure}

\begin{deluxetable*}{lccccc}
\tabletypesize{\footnotesize}
\tablewidth{0pc}
\tablecaption{O32 Subsamples}
\tablehead{
\colhead{O32 bin} &
\colhead{$N_{\rm comp}$\tablenotemark{a}} &
\colhead{$\langle z\rangle$\tablenotemark{b}} &
\colhead{$\langle {\rm O32}\rangle$\tablenotemark{c}} &
\colhead{$\langle \rsii\rangle$\tablenotemark{d}} &
\colhead{$\langle \logq\rangle$\tablenotemark{e}}}
\startdata
all & 48 & $4.373$ & $3.457\pm0.074$ & $1.355\pm0.100$ ($0.303$) & $54.50\pm0.01$ ($0.09$) \\
low  & 24 & $4.063$ & $1.978\pm0.057$ & $1.910\pm0.116$ ($0.312$) & $54.51\pm0.02$ ($0.15$)\\
high & 24 & $4.683$ & $6.055\pm0.174$ & $0.987\pm0.123$ ($0.204$) & $54.50\pm0.02$ ($0.10$)\\
\enddata
\tablenotetext{a}{Number of galaxies with spectral coverage of $\sii$, used to construct the composite spectra.}
\tablenotetext{b}{Mean redshift of galaxies in this bin.}
\tablenotetext{c}{Mean and uncertainty in mean O32 ratio for galaxies in this bin.}
\tablenotetext{d}{Mean and uncertainty in mean $\rsii$ for galaxies in this bin.  Numbers in parentheses include sample variance.}
\tablenotetext{e}{Mean and uncertainty in mean $\logq$ for galaxies in this bin.  Numbers in parentheses include sample variance.}
\label{tab:rsiio32}
\end{deluxetable*}

Owing to the relatively low $S/N$ of the $\sii$ doublet in the spectra
of individual galaxies, we primarily limit our discussion to the
results obtained from composite spectra, though for reference
individual measurements are included in
Figure~\ref{fig:densityq}---these individual measurements are obtained
for the 12 (of 48) objects in the sample where both lines of the
$\sii$ doublet are detected at $\ge 3\sigma$.  To maximize the
signal-to-noise of the composite spectrum, and in light of the absence
of strong evolution in the ionization conditions in the ISM between
$z\sim 2$ and $z\sim 6$ \citep{sanders23c}, composite spectra were
constructed using galaxies over the full range of redshift ($z\sim
2.7-6.3$) of the sample.  

Based on the composite spectrum of all 48 galaxies, we find
$\langle\rsii\rangle = 1.355\pm0.100$, corresponding to $\langle
n_e\rangle = 64_{-63}^{+84}$\,cm$^{-3}$ (Table~\ref{tab:rsiio32}).
This is somewhat lower than the $n\simeq 250$\,cm$^{-2}$ derived for
more massive star-forming galaxies at $z\sim 2.3$ (with a median
stellar mass of $M_\ast \simeq 10^{10}$\,$M_\odot$) from the MOSFIRE
Deep Evolution Field (MOSDEF) survey \citep{sanders16b}.  The mean
electron density found here is also lower than the $n_e\ga
300$\,cm$^{-3}$ derived for individual galaxies at $z=4.0-8.7$ from
the line-spread-function deconvolution of $\oii$\,$\lambda\lambda
3727, 3730$ by \citet{isobe23}, where $\oii$ is not resolved for most
of the objects in their sample.  Five of the galaxies in their
analysis are also in our sample.  At face value, the sample-averaged
lower $\langle n_e\rangle$ found here may be due to higher fraction of
galaxies in our sample with lower $n_e$.  A larger sample of galaxies
with $n_e$ measured from the same feature (i.e., $\sii$ or $\oii$)
should help to clarify the source of this difference.

Composite spectra were also constructed in two equal-number bins of
O32.  The $\sii$ doublet in the composite spectra for the low- and
high-O32 bins is shown in Figure~\ref{fig:specplot}.  Properties of
these bins, and the mean O32 and $\rsii$ measured from the composite
spectra, are listed in Table~\ref{tab:rsiio32}.  The mean O32 and
$\rsii$ measured from the composite spectra are shown in
Figure~\ref{fig:densityq}.  The data reveal a potential correlation
between $\rsii$ and O32, or an inverse correlation between $n_e$ and
O32.  Specifically, the difference in $\langle\rsii\rangle$ between
the two O32 bins is significant at the $5.5\sigma$ level.  The
differences in $\langle\rsii\rangle$ between the O32 bins suggest that
$\langle n_e\rangle$ increases with O32: in this case, galaxies with
$\langle{\rm O32}\rangle \simeq 6$ have $\langle n_e\rangle \simeq
500$\,cm$^{-3}$, a factor of at least $\simeq 5\times$ larger than the
$\langle n_e\rangle $ inferred for galaxies with $\langle{\rm
  O32}\rangle \simeq 2$.  A similar trend of increasing $n_e$ with
O32, and hence $U$, has been noted locally \citep{bian16, jiang19} and
at $z\sim 2-3$ \citep{shirazi14, reddy23b}, and inferred at $z\sim
1.1-2.3$ \citep{papovich22}.  Our results suggest that such a trend
continues unabated up to $z\sim 6.3$.

While these results imply a significant correlation between $n_e$ and
O32, an accounting of sample variance (Section~\ref{sec:data}) implies
a rather large scatter in this correlation.  In other words, there are
some realizations of the sample---obtained by constructing composites
of randomly selected objects with replacement---that result in
$\langle\rsii\rangle$ that are not significantly different between the
low and high O32 bins.  Likewise, there are some sample realizations
where the difference in $\langle\rsii\rangle$ between the low and high
O32 bins is highly significant.  The combined effects of measurement
uncertainty and sample variance imply a $\simeq 2.5\sigma$ difference
in $\langle n_e\rangle$ between the low- and high-O32 populations.

We now turn to the variation of $Q$ with O32 (bottom panel of
Figure~\ref{fig:densityq}).  A Spearman correlation test on the
individual measurements for all 48 galaxies in the O32 sample implies
a $p=0.67$ probability that the relationship between $\logq$ and O32
occurs by random chance if the two parameters are uncorrelated with
each other.  The mean $\logq$ measured in two equal-number bins of O32
are listed in Table~\ref{tab:rsiio32} and shown by the large diamonds
in the bottom panel of Figure~\ref{fig:densityq}.  We find no
significant difference in $\langle\logq\rangle$ between the low- and
high-O32 populations.

To summarize, we find evidence for increased $n_e$ among galaxies with
higher O32 (though with large scatter), suggesting that variations in
$n_e$ may play a role in the elevated ionization parameters inferred
for high-redshift galaxies.  On the other hand, we find no significant
correlation between $Q$ and O32.  These results are discussed further
in the next section.

\section{Results and Discussion}
\label{sec:discussion}

In the previous section, we presented evidence suggesting that
variations in $n_e$ may partly explain the high O32 (and $U$) inferred
for high-redshift galaxies, while $Q$ does not appear to be an
important factor in driving high O32 within the sample.  Here, we
examine these possibilities in the context of radiation- and
density-bounded nebulae (Sections~\ref{sec:radiationneb} and
\ref{sec:densityneb}).  These results are discussed in the context of
previous studies that have addressed the roles of metallicity
(Section~\ref{sec:metallicity}) and SFR surface density
(Section~\ref{sec:sigmasfr}) in modulating $U$ at high redshift.

\subsection{Expectations for a Radiation-bounded Nebula}
\label{sec:radiationneb}

For a radiation-bounded nebula, we expect $U\propto Q^{1/3}n_e^{1/3}$
(Equation~\ref{eq:mainequation}).  In this case, the increase in
$\langle n_e\rangle$ from the low-density regime ($n_e \sim
1-10$\,cm$^{-3}$) for the low-O32 population to $n_e \simeq
500$\,cm$^{-3}$ for the high-O32 population is expected to yield a
factor of $\simeq 4-8$ increase in $\log U$, or a similar factor
increase in O32.  This expected increase in O32 due to changes in
$n_e$ is more than sufficient to account for the factor of $\approx 3$
increase in $\langle{\rm O32}\rangle$ between the low- and high-O32
populations (Table~\ref{tab:rsiio32}).  This also suggests that the
nebulae are reasonably approximated as radiation bounded; $U$ will
scale {\em inversely} with gas density for a density-bounded nebula
(Section~\ref{sec:densityneb}), which is inconsistent with the
inferred {\em increase} in $U$ with $n_e$ (Figure~\ref{fig:densityq}).
As there is no significant change in $\langle Q\rangle$ between the
two O32 bins, the increase in O32 within these redshift bins is likely
to be driven at least in part by an increase in $n_e$.

While $Q$ does not appear to be a significant factor in driving O32
within the sample, this does not preclude the possibility of a
connection between $Q$ and $U$.  In particular, \citet{reddy23b}
suggest that the combined effects of an increasing $n_e$ and SFR (or
$Q$) may be responsible for explaining the evolution towards higher
$U$ from $z\sim 0$ to $z\sim 2$ at a fixed stellar mass (see also
\citealt{brinchmann08, shirazi14, bian16, kaasinen17, papovich22,
  reddy23b}).  The lack of a significant correlation between $Q$ and
O32 found here may simply reflect the limited dynamic range in SFR
probed by our sample (but see Section~\ref{sec:densityneb}).  The
factor(s) driving the changes in $n_e$ and $Q$ are discussed in
Section~\ref{sec:sigmasfr}.

\subsection{Expectations for a Density-bounded Nebula}
\label{sec:densityneb}

Rest-frame far-UV spectroscopic observations imply that typical
galaxies at $z\sim 2-3$ have an ISM configuration where most
sightlines are radiation bounded, while the remaining fraction of
sightlines coincide with low-column-density or ionized
(density-bounded) channels \citep{reddy16b, steidel18, reddy22}.  A
low covering fraction of radiation-bounded sightlines will result in
higher O32 even for a fixed intensity or hardness of the input
(stellar) ionizing spectrum (e.g., \citealt{giammanco05, brinchmann08,
  nakajima13}).  There are at least three consequences of such a
configuration.  First, there will be a non-zero fraction of ionizing
photons that escapes the ISM.  Second, O32 will overestimate $U$.  For
example, an $\hii$ region with a $50\%$ escape fraction of ionizing
photons ($f_{\rm esc}$) would result in an order-of-magnitude
overestimate of $U$ compared to a region with zero escape fraction
\citep{giammanco05}.  Third, in the limit where the covering fraction
of radiation-bounded sightlines falls to zero (i.e., a purely
density-bounded nebula), $R_{\rm S}$ will simply represent the
physical extent of the cloud and it will not depend on $n_e$.  As a
result, $U$ will be inversely proportional to $n_e$
(Equation~\ref{eq:1}).  Of course, real galaxies are unlikely to be
purely density bounded and, as a consequence, one would expect $U$ to
transition from a $n_e^{1/3}$ dependence to a close to $n_e^{-1}$
dependence as $f_{\rm esc}$ increases from $f_{\rm esc}=0$ to close
to unity.

\begin{figure}
  \epsscale{1.2}
  \plotone{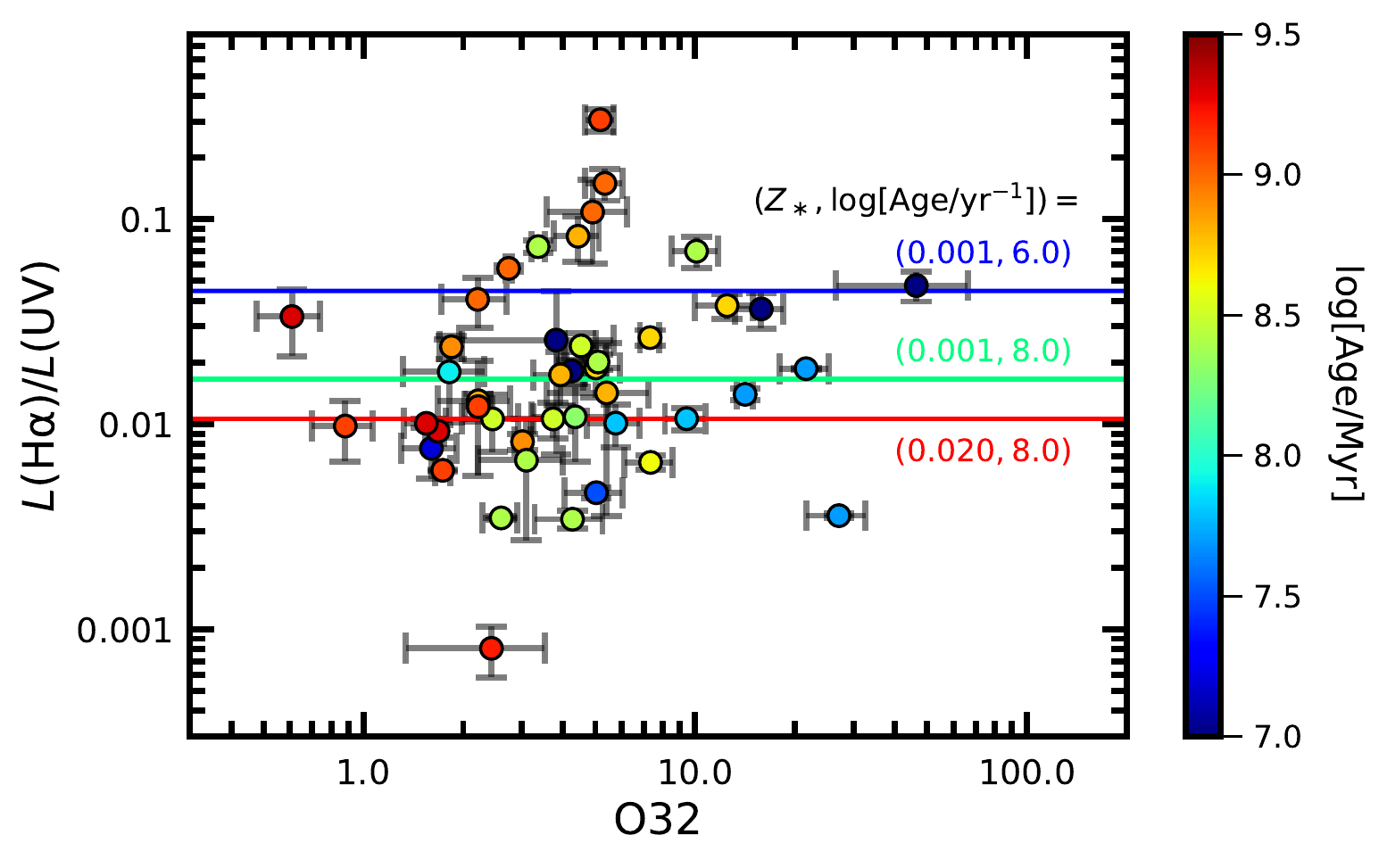}
    \caption{Variation of $\lha/\luv$ with O32.  Points are color
      coded by the best-fit stellar population age derived from SED
      fitting (Section~\ref{sec:data}).  Horizontal lines indicate the
      $\lha/\luv$ expected for the BPASS v2.2.1 binary
      constant-star-formation models for the indicated $Z_\ast$ and
      age.}
    \label{fig:hauv}
\end{figure}

It is prudent to investigate whether elevated $U$ (or O32) may be tied
to a high $f_{\rm esc}$.  An increase in $f_{\rm esc}$ suppresses
$\hi$ recombination line fluxes relative to the non-ionizing UV
continuum.  If high escape fractions are responsible for driving
elevated O32 ratios, one might expect the ratio of the
(dust-corrected) $\ha$ and UV luminosities, $\lha/\luv$, to
anti-correlate with O32.  Figure~\ref{fig:hauv} shows the variation of
$\lha/\luv$ with O32 for the 43 objects in the sample with
well-constrained SED fits around $1500$\,\AA, and thus robust $\luv$
determinations.  The data do not indicate that galaxies with higher
O32, in particular those with ${\rm O32}>10$, have systematically
lower $\lha/\luv$.

However, it is important to keep in mind that $\lha/\luv$ is also
sensitive to the stellar metallicity ($Z_\ast$), IMF, and age of the
stellar population.  The variation in $\lha/\luv$ arising from stellar
population differences (e.g., different $Z_\ast$ and age) for the
Binary Population and Stellar Synthesis (BPASS; \citealt{eldridge17,
  stanway18}) v2.2.1 constant-star-formation (CSF) models with an
upper-mass cutoff of the IMF of $100$\,$M_\odot$ is illustrated in
Figure~\ref{fig:hauv}.  Galaxies with lower stellar metallicities and
younger ages will exhibit higher intrinsic $\lha/\luv$.  Indeed, the
stellar population ages derived from fitting the broadband SEDs of
galaxies in the sample (Section~\ref{sec:data}; \citealt{shapley23a})
are systematically younger for those with higher O32, as indicated by
the color-coding in Figure~\ref{fig:hauv}.  If galaxies with high O32
are preferentially undergoing a burst of star formation, and such a
burst is accompanied by a high $f_{\rm esc}$ due to an increase in
ISM porosity related to feedback (e.g., \citealt{trebitsch17, kimm19,
  ma20, kakiichi21}), then the deficit of $\lha$ relative to $\luv$
expected for a non-zero escape fraction may be masked by the higher
intrinsic $\lha$ relative to $\luv$ for a bursty (or young)
low-metallicity galaxy.  

We cannot conclusively rule out this possibility with the data at
hand.  An analysis of the gas covering fraction of galaxies with high
O32 may elucidate the role of ionizing photon leakage in such
galaxies.  Nevertheless, invoking a very high $f_{\rm esc}$ to explain
the high O32 observed for some galaxies \citep{giammanco05,
  brinchmann08, nakajima13} necessarily implies that a large fraction
of the gas must be ionized, which in turn would have the effect of
suppressing star formation and the production of ionizing photons.  A
large representative sample of galaxies at $z\ga 2.7$ may be required
for obtaining statistically robust constraints on the O32 distribution
and the duty cycle of a (presumably) transient and short phase of
elevated O32 and high $f_{\rm esc}$ (e.g., \citealt{naidu22}).

Note that dust-corrected $\lha$ will underestimate $Q$ if there is a
non-zero escape fraction of ionizing photons, or if there is
significant dust extinction of ionizing photons within the $\hii$
regions (Section~\ref{sec:dust}).  It is possible that correcting for
these effects may yield a more significant correlation between $Q$ and
O32 (Figure~\ref{fig:densityq}), {\em if} the average escape fraction
is increasing with O32.  However, the expectation of such a
correlation is complicated by the large scatter in $\logq$ that may be
driven by stellar population differences from galaxy to galaxy.  The
upshot is that it is difficult to use the $\lha/\luv$ ratio alone to
distinguish galaxies that may be leaking a significant fraction of
ionizing photons, and one should turn to other constraints in this
regard (e.g., escape fraction and kinematics of Ly$\alpha$, gas
covering fraction, etc.).

\subsection{Metallicity Considerations}
\label{sec:metallicity}

The well-established anti-correlation between $U$ and $Z_{\rm neb}$
for local star-forming galaxies (e.g., \citealt{dopita86, dopita06,
  perez14}) is typically interpreted as a reflection of the harder and
more intense ionizing spectra associated with lower-metallicity
massive stars \citep{dopita06, leitherer14}.  If so, one would expect
$U$ to be driven primarily by changes in $Z_{\rm neb}$.  However, the
hardness of the ionizing spectrum (which $U$ is sensitive to) is
expected to correlate more directly with $Z_\ast$ than with $Z_{\rm
  neb}$, since the former is responsible for regulating stellar
opacity and the absorption of ionizing photons by stellar winds
\citep{dopita06}.  Hence, investigating variations in $U$ with
$Z_\ast$ presents a potentially more direct route to assessing the
role of metallicity in modulating $U$ \citep{strom18,reddy23b},
particularly in cases where the Fe-sensitive $Z_\ast$ may differ from
the O-sensitive $Z_{\rm neb}$ as is the case for the $\alpha$-enhanced
stellar populations typical of $z\ga 2$ galaxies \citep{steidel16,
  cullen19, topping20a, topping20b, cullen21, runco21, reddy22}.  As
we discuss below, the relationship between $U$ and $Z_{\rm neb}$ is
still relevant in determining the effects dust on the intensity and
hardness of the ionizing radiation field.  The expected dependence of
$U$ on $Z_\ast$ is discussed below, but first we address the variation
of $U$ with $Z_{\rm neb}$ within the sample.

The photoionization modeling used to simultaneously constrain $\log U$
and $\log[Z_{\rm neb}/Z_\odot]$ is described in
Section~\ref{sec:photoionization}.  The effect of $Z_{\rm neb}$ and
$Z_\ast$ on $U$ is discussed in Section~\ref{sec:xiion}.  The
possibility that metallicity may affect $n_e$ is addressed in
Section~\ref{sec:metallicitydensity}, and we also consider the
diminution and softening of the ionizing radiation field in dusty
$\hii$ regions in Section~\ref{sec:dust}.
Section~\ref{sec:modelsarewrong} addresses whether the stellar
population synthesis models incorrectly predict how rapidly the
hydrogen-ionizing spectrum hardens with decreasing metallicity below
$Z_\ast \simeq 0.001$.

\subsubsection{Photoionization Modeling}
\label{sec:photoionization}

\begin{figure*}
  \epsscale{1.2}
  \plotone{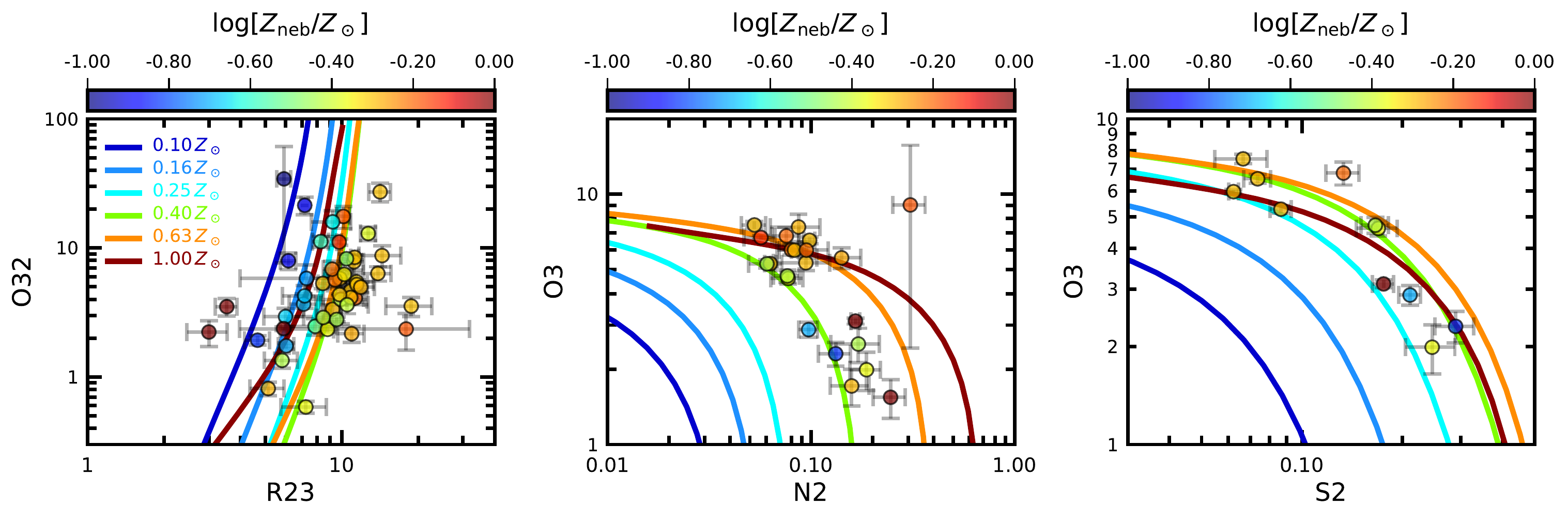}
    \caption{O32 vs. R23 (left), and the N2 and S2 BPT diagrams
      (middle and right, respectively).  Photoionization modeling
      predictions for how the line ratios vary with $\log U$ at fixed
      values of $Z_{\rm neb}$ are indicated by the colored curves.
      The modeling results shown here assume the fiducial BPASS model
      (i.e., a fixed ionizing spectrum), and we note that the curves
      are relatively insensitive to the assumed age and metallicity at
      $Z_\ast \la 0.002$ (see text).  Data points for galaxies where
      the relevant lines are detected with ${\rm S/N}\ge 3$ are color
      coded by the best-fit $\log[Z_{\rm neb}/Z_\odot]$ found from
      photoionization modeling of all the significantly-detected
      lines.}
    \label{fig:zneb}
\end{figure*}

\begin{figure}
  \epsscale{1.1}
  \plotone{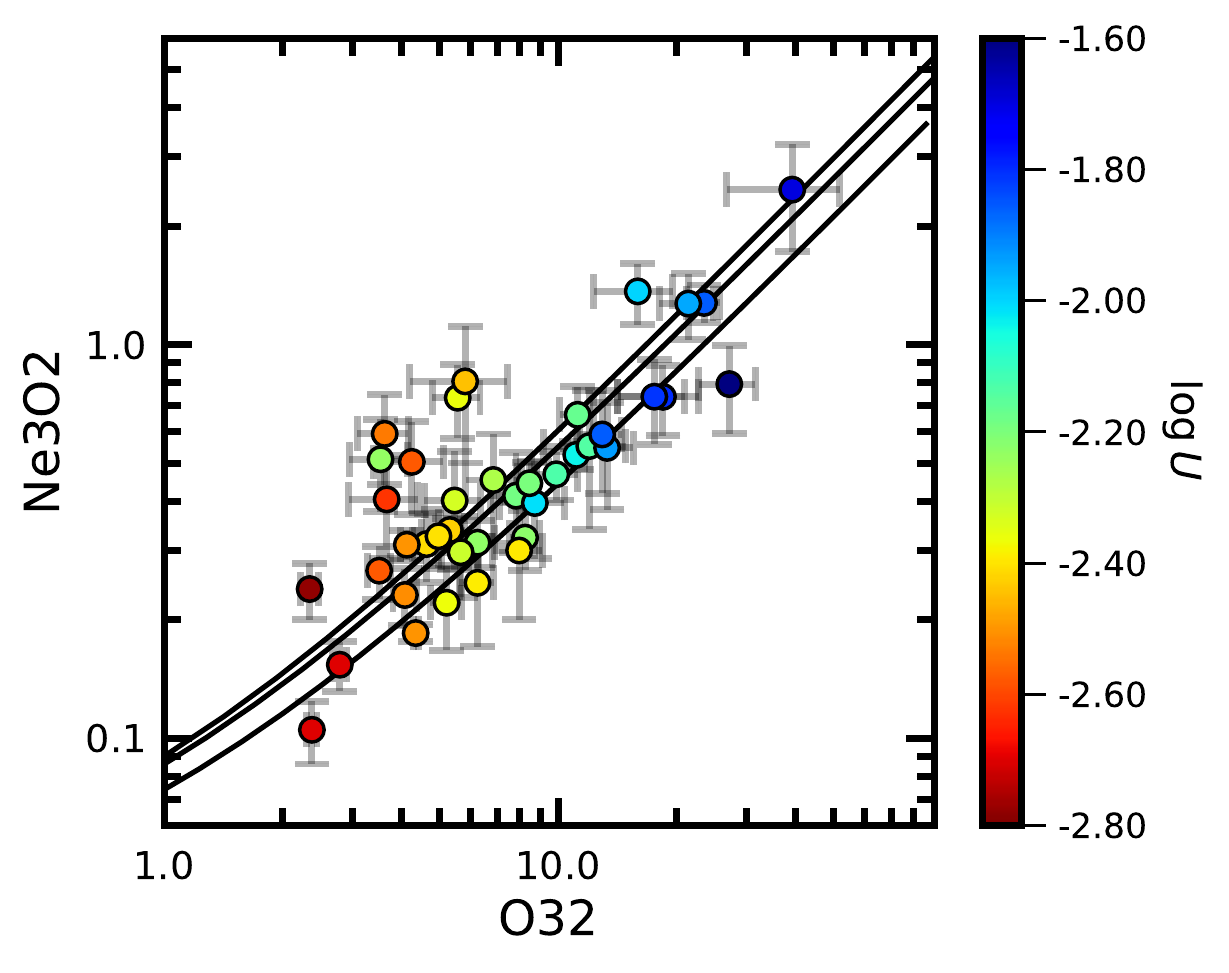}
    \caption{Ne3O2 vs. O32 for the 40 objects in the sample where all
      the relevant lines are detected at $\ge 3\sigma$ (circles),
      color coded by the inferred $\log U$.  The model predictions for
      how the line ratios vary with $\log U$ are denoted by the three
      black curves for three values of oxygen abundance: $\log[Z_{\rm
          neb}/Z_\odot] = -1.0, -0.5, 0.0$.}
    \label{fig:logu}
\end{figure}

Photoionization modeling was used to simultaneously determine $Z_{\rm
  neb}$ and $\log U$ using all available rest-frame optical emission
lines, including $\oii$\,$\lambda\lambda 3727,3730$,
$\neiii$\,$\lambda 3870$, $\hd$, $\hg$, $\hb$,
$\oiii$\,$\lambda\lambda 4960, 5008$, $\oi$\,$\lambda 6302$, $\ha$,
$\nii$\,$\lambda 6585$, and $\sii$\,$\lambda 6718, 6733$---the $\hi$
recombination lines aid in determining the ``best-fit'' value of
nebular reddening, which is then used to correct all other lines for
dust attenuation.  The CLOUDY v17.02 radiative transfer code
\citep{ferland17} was used for the photoionization modeling, where
$\log[Z_{\rm neb}/Z_\odot]$ was allowed to vary in the range $-2.0\le
\log[Z_{\rm neb}/Z_\odot] \le 0.0$ in increments of 0.1\,dex, and
$\log U$ was allowed to vary in the range $-3.5\le \log U\le -1.0$ in
increments of 0.1\,dex.  We further assumed an input ionizing spectrum
corresponding to the BPASS v2.2.1 binary CSF model with $Z_\ast =
0.001$, $\log[{\rm Age/Myr}] = 8.0$, and an upper-mass cutoff of the
IMF of $100$\,$M_\odot$, values typical of those inferred based on
fitting the BPASS models to the rest-frame far-UV spectra of
star-forming galaxies at $z\sim 2-3$ \citep{steidel16, topping20a,
  reddy22}.  In this subsequent discussion, we refer to this as the
``fiducial'' BPASS model.  The hydrogen density was fixed to $n_{\rm
  H} = 250$\,cm$^{-3}$, similar to the average density inferred in
Section~\ref{sec:neq} and in previous investigations of $z\sim 2$
galaxies (e.g., \citealt{sanders16a, topping20b, reddy23b}).  Note
that while we have argued that density plays a role in modulating $U$
(Section~\ref{sec:neq}), the $\log U$ inferred from photoionization
modeling is relatively insensitive to the choice of $n_{\rm H}$
because the {\em translation} between O32 and $U$ is insensitive to
$n_{\rm H}$ (see discussion in \citealt{reddy23b}).  In other words,
for a fixed set of measured line ratios, the inferred $U$ is
relatively insensitive to the choice of $n_{\rm H}$, or, for that
matter, the input ionizing spectrum over the range of stellar
metallicity ($Z_\ast \la 0.002$), age ($\log[{\rm Age/Myr}] =
7.0-8.0$), and $n_{\rm H}$ ($n_{\rm H} \simeq 10-500$\,cm$^{-3}$).
Anywhere from 6 to 11 lines with ${\rm S/N}\ge 3$ were fit
simultaneously for each object to constrain $Z_{\rm neb}$ and $U$.
Uncertainties were determined from the dispersion of $Z_{\rm neb}$ and
$U$ obtained by randomly perturbing the line fluxes by their errors
many times and refitting the photoionization models to these
realizations.

The efficacy of the models in reproducing line ratios sensitive to
oxygen abundance ($Z_{\rm neb}$) is illustrated in
Figure~\ref{fig:zneb}.  Measured line ratios for galaxies where the
relevant lines used to construct the ratios are detected with ${\rm
  S/N} \ge 3$ are color-coded by the best-fit $\log[Z_{\rm
    neb}/Z_\odot]$ obtained from the photoionization modeling.  Of the
48 galaxies in the sample, 26 and 36 have upper limits on N2 and S2,
respectively, and are therefore not shown in Figure~\ref{fig:zneb}.
Naturally, these galaxies tend to have lower oxygen abundances based
on their distribution of O32 and R23.  The best-fit values can be
compared with the model predictions of how the line ratios vary with
$\log U$ at a fixed $Z_{\rm neb}$ (colored curves, where each curve
represents a sequence in $\log U$ at a fixed $Z_{\rm neb}$).  In
almost all cases, the models are able to reproduce the measured line
ratios within $3\sigma$.  There are two objects (CEERS ID 2422 and
2693) where N2 and S2 indicate a higher $\log[Z_{\rm neb}/Z_\odot]\ga
-0.4$ than inferred from R23, $\log[Z_{\rm neb}/Z_\odot]\la -0.8$.
Additionally, the two objects (CEERS ID 2089 and 2514) with the
highest inferred $Z_{\rm neb}$ in the sample (close to solar) are
offset to lower R23 than the $Z_{\rm neb} = Z_\odot$ curve would
predict for their O32---and similarly offset toward lower N2 than the
$Z_{\rm neb} = Z_\odot$ curve would predict for their O3.  A higher
($\ga$solar) stellar metallicity and/or super-solar O/H
may more appropriately describe these two galaxies.  At any rate, the
presence of the aforementioned objects in the sample does not alter
the subsequent discussion or conclusions regarding the scatter in
$\log U$ at a fixed $Z_{\rm neb}$.

Particularly tight constraints on $\log U$, with little dependence on
$Z_{\rm neb}$, are afforded by the combination of O32 and Ne3O2 (e.g.,
\citealt{nagao06, perez07, levesque14, steidel16, strom17, jeong20,
  shapley23b}).  The latter is relatively insensitive to dust
corrections owing to the wavelength proximity of $\neiii$\,$\lambda
3870$ and $\oii$\,$\lambda\lambda 3727, 3730$.  Figure~\ref{fig:logu}
displays the Ne3O2 and O32 measurements for the 40 galaxies in the
sample where the line ratios could be measured (i.e., 8 of the 48
objects in the sample did not have $\ge 3\sigma$ detections of
$\neiii$).  Model predictions for Ne3O2 versus O32 are shown for the
fiducial BPASS model and three values of $\log[Z_{\rm neb}/Z_\odot] =
-1.0, -0.5, 0.0$ (black curves).  The average Ne3O2 measured in the
low- and high-O32 bins are also indicated in the figure with the large
diamonds.  The measured line ratios are in excellent agreement with
the model predictions and yield stringent constraints on $\log U$,
with a typical uncertainty of $\simeq 0.1$\,dex.

\begin{figure}
  \epsscale{1.2}
  \plotone{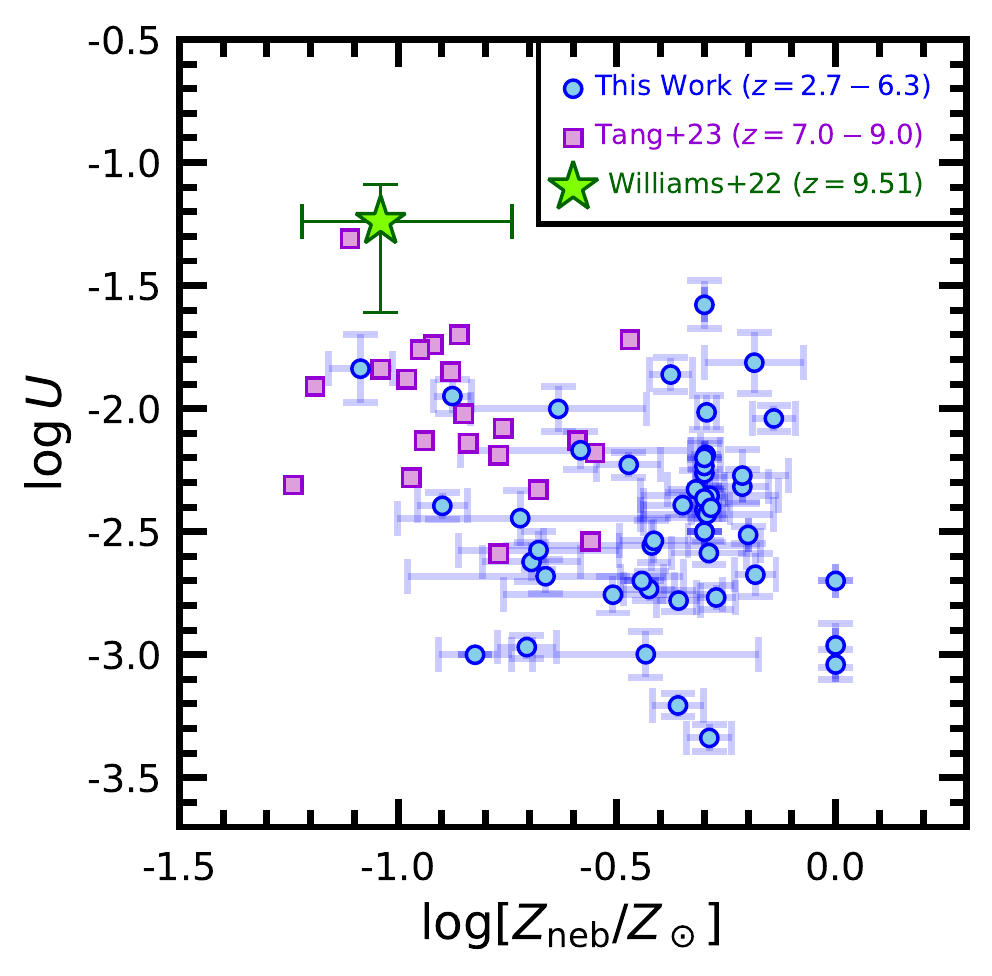}
    \caption{$\log U$ vs. $\log[Z_{\rm neb}/Z_\odot]$ for the 48
      galaxies in the sample (blue points).  For comparison, the data
      from \citet{tang23} for $z=7.0-9.0$ galaxies (purple squares, error bars
      suppressed for clarity), and a lensed $z=9.51$ galaxy from
      \citet{williams22} (green star), are also shown.}
    \label{fig:uvsz}
\end{figure}

\subsubsection{Variation of $U$ with $Z_{\rm neb}$ and $Z_{\ast}$}
\label{sec:xiion}

Figure~\ref{fig:uvsz} shows the distribution of $\log U$ and
$\log[Z_{\rm neb}/Z_\odot]$ for the sample.  A Spearman test indicates
a high probability ($p=0.95$) of a null correlation between $\log U$
and $\log[Z_{\rm neb}/Z_\odot]$.  Deeper observations of multiple
abundance-sensitive line ratios (including those involving auroral
lines such as $\oiii$\,$\lambda 4364$) should allow us to obtain more
stringent constraints on $Z_{\rm neb}$ and reevaluate the significance
of any correlation between $\log U$ and $\log[Z_{\rm neb}/Z_\odot]$.
Note that the $\simeq 1$\,dex order of magnitude dynamic range in
$\log[Z_{\rm neb}/Z_\odot]$ probed by the sample may be insufficient
to reveal the expected anti-correlation between $U$ and $Z_{\rm neb}$
given the large scatter between these quantities.  For example,
\citet{sanders20} and \citet{topping20b} do find a significant
anti-correlation between these parameters when considering a larger
dynamic range in $Z_{\rm neb}$.  For comparison, data from
\citet{tang23} for $z=7-9$ galaxies in CEERS, along with the $z=9.51$
lensed galaxy from \citet{williams22}, are shown in
Figure~\ref{fig:uvsz}.  The combination of these datasets suggests
that galaxies with lower $Z_{\rm neb}$ have higher $U$, at least on
average, similar to the behavior seen locally \citep{perez14}.

In any case, it is clear that at a fixed $\log[Z_{\rm neb}/Z_\odot]
\simeq -0.3$, there is a $\ga 1.5$ order of magnitude variation in
$\log U$ within our sample, which ranges from $\log U \simeq -3.3$ to
$-1.6$, and is much larger than the typical $0.1-0.2$\,dex measurement
uncertainties in $\log U$.  This large variation of $\log U$ at a
fixed $Z_{\rm neb}$ has also been seen in lower-redshift ($z\sim 2$)
galaxies, albeit over a smaller dynamic range in $Z_{\rm neb}$ (see
Figure~10 of \citealt{topping20b}).
The large spread in $U$ at a fixed $Z_{\rm neb}$ implies that factors
other than $Z_{\rm neb}$ are driving the variation in $\log U$ within
the sample.

As mentioned above, the metallicity dependence of $U$ can be more
directly examined with estimates of $Z_\ast$.  Using the BPASS models
discussed above, \citet{reddy23b} evaluated the expected impact of
changes in $Z_\ast$ on $U$.  For example, $\logxi = 25.39$ for the
fiducial BPASS model.  Given the evolution of decreasing metallicity
with redshift (e.g., at a fixed mass), one would expect the $z>2.7$
galaxies in our sample to have stellar metallicities similar to or
less than $Z_\ast \simeq 0.001$.  Keeping all other parameters fixed
and lowering the stellar metallicity to an ``extreme'' value of
$Z_\ast = 10^{-4}$ yields $\logxi = 25.45$, implying $\Delta\logxi =
0.06$ relative to the fiducial model (and hence a similar change in
$\log Q$; Section~\ref{sec:neq}).  For a radiation-bounded nebula
where $U\propto Q^{1/3}$, these changes in $Z_\ast$ imply a
corresponding change in $\log U$ of $0.02$\,dex, much smaller than the
observed $1.5$\,dex variation in $U$ among galaxies in the present
sample.  Thus, over the range of stellar metallicities expected for
$z\ga 2$ galaxies ($Z_\ast \la 0.002$), and hence over the range of
associated ionizing spectral shapes, the models predict that
metallicity alone is insufficient to explain the observed variations
in $U$.

Of course, larger modulations in $U$ are possible when combining
variations in $Z_\ast$ with a change in stellar population age and/or
the IMF.  For example, the BPASS binary models predict $\logxi =
25.93$ for a higher upper-mass cutoff of the IMF of $300$\,$M_\odot$,
$Z_\ast = 10^{-4}$, and $\log[{\rm Age/Myr}] = 6.0$ (i.e., an
extremely short burst of star formation).  Hence, $\Delta\logxi =
0.54$ relative to the fiducial model.  Even in this case, the impact
of such a change on $\log U$ is modest; i.e., $\Delta\log U =
0.18$\,dex, again quite small compared to the spread in $\log U$
inferred for the sample.  The bottom line of this discussion is that
it may be difficult to fully explain the variation in $U$ within the
sample with the combined effects of stellar metallicity and other
stellar population differences (e.g., age, IMF) alone.  In
Section~\ref{sec:modelsarewrong}, we consider the possibility that the
models may not correctly predict the ionizing spectrum.

\subsubsection{Effect of Metallicity on $n_e$}
\label{sec:metallicitydensity}

We have just considered the impact of metallicity on $\xi_{\rm ion}$,
and hence $Q$.  The effects of metallicity on $U$ may be more
pronounced than indicated in that previous discussion if $n_e$ also
depends on metallicity.  Energy deposition into $\hii$ regions from
stellar winds, for example, may result in higher $\hii$ region
pressures and densities (e.g., \citealt{groves08, krumholz09,
  kaasinen17, jiang19, davies21}), and it has been well established
that wind mass-loss rates are sensitive to stellar metallicity (e.g.,
\citealt{kudritzki00, vink01, brott11, langer12, vink22}).  However,
stellar winds become {\em weaker}---i.e., wind mass-loss rates
decrease---with decreasing stellar metallicity, opposite of the
direction needed to explain an anti-correlation between $U$ and
$Z_\ast$ through stellar feedback alone.  \citet{davies21} explore in
detail a few of the physical drivers that may be responsible for the
higher $n_e$ observed in high-redshift galaxies, including molecular
cloud density, stellar feedback, ambient medium density, and dynamical
evolution of $\hii$ regions.  Assessing these possibilities and how
they may relate to the stellar or $\hii$-region metallicity will be
necessary for a full accounting of the effect of metallicity on $n_e$
and the volume filling fraction, $\epsilon$.  Regardless of the manner
in which metallicity regulates $n_e$ and $\epsilon$, if at all, the
net impact on $\log U$ appears to be inconsequential given the large
scatter between $\log U$ and $\log[Z_{\rm neb}/Z_\odot]$.


\subsubsection{Internal Dust Attenuation in $\hii$ Regions}
\label{sec:dust}

While $Z_\ast$ may be more relevant for assessing the hardness of the
emergent ionizing stellar spectrum, the impact of dust on the
effective ionizing spectrum seen by the gas cloud is likely connected
to $Z_{\rm neb}$.  In particular, it is well known that the
dust-to-gas ratio correlates with oxygen abundance \citep{draine07b,
  ruyer14}.  Dust within $\hii$ regions will reduce the intensity of
the ionizing spectrum, soften the ionizing spectrum due to the
wavelength dependence of dust extinction (e.g., \citealt{inoue01a}),
and increase the radiation pressure in the $\hii$ region through
dynamic coupling of the dust and gas \citep{draine11, yeh12, ali21}.
Consequently, the ionizing radiation field permeating the gas cloud
will be softer in higher-metallicity regions, owing both to a higher
stellar opacity and dust within the $\hii$ region, thus resulting in
lower $Q_{\rm eff}$ (i.e., the effective ionizing photon rate
including dust absorption of ionizing photons) and lower $U$.  On the
other hand, radiation pressure can increase the inhomogeneity of the
density profile as gas is piled up near the ionization front,
potentially increasing $\epsilon$ (e.g., \citealt{krumholz09, ali21}).

The balance between these various effects and their impact on $U$ is
beyond the scope of this work.  However, the typical densities
inferred for high-redshift $\hii$ regions ($n_e\ga 100$\,cm$^{-3}$)
are sufficiently high that photoelectric absorption is expected to
dominate over dust absorption even assuming a Milky Way dust-to-gas
ratio of $\simeq 1\%$.  Furthermore, significant dust absorption of
ionizing photons does not appear to be supported by the
self-consistent modeling of the non-ionizing UV continuum and
rest-frame optical emission line ratios which are sensitive to the
shape of the permeating radiation field in typical (moderately dusty)
star-forming galaxies at $z\sim 2-3$ \citep{steidel16, topping20a,
  reddy22}.  Finally, as with the discussion in the previous section,
if $Z_{\rm neb}$ is indicative of the dust-to-gas ratio in the $\hii$
regions, then dust does not appear to be the main driver of variations
in $U$ given the large scatter between $U$ and $Z_{\rm neb}$ within
our sample.

\subsubsection{Uncertainties in the Ionizing Spectra of Low-Metallicity Stars}
\label{sec:modelsarewrong}

A relevant point of inquiry concerns the validity of the hydrogen
ionizing spectrum (between 1 and 4 Rydbergs) predicted by the BPASS
models, particularly at lower metallicity where direct empirical
constraints on the shape and intensity of the ionizing spectrum are
lacking.  Metallicity could play a more prominent role in shaping $U$
if the ionizing spectrum hardens or intensifies more rapidly with
decreasing metallicity than what the models predict.  A subsequent
increase in $\lha/\luv$ with decreasing metallicity may be difficult
to discern from the observations given other variations in the
properties of the stellar population or escape fraction of ionizing
photons (Section~\ref{sec:densityneb}).  

However, we can look to lower-redshift ($z\sim 2-3$) galaxies for
guidance in this regard.  Specifically, typical ($\simeq L^\ast$ or
$M^\ast$) star-forming galaxies at these redshifts can be
approximately characterized with smoothly-varying star-formation
histories (e.g., \citealt{papovich11, reddy12b, pacifici15}),
particularly when averaged over a statistical sample of galaxies.
Rest-frame UV spectra have also been used to characterize the stellar
metallicities of these galaxies (e.g., \citealt{steidel16, cullen19,
  topping20a, topping20b, reddy22}).  In addition, the far-UV spectra
yield constraints on the gas covering fraction of optically-thick
$\hi$, which has been shown to anti-correlate strongly with ionizing
(and $\lya$) escape fraction at these redshifts \citep{steidel18,
  reddy22}.  Finally, a large body of work has focused on constraining
the stellar and nebular dust attenuation in typical $z\sim 2$ galaxies
(e.g., \citealt{reddy06a, daddi07a, reddy10, buat12, pannella15,
  shivaei15a, debarros16, shivaei16, reddy18a, reddy20, shivaei20a};
and references therein).

Employing the BPASS models and the best available constraints on
stellar metallicity, star-formation history, age, dust attenuation,
and ionizing escape fraction, \citet{reddy22} found a good agreement
between ionization-rate-based SFRs ($\sfrha$) and non-ionizing UV
continuum-based SFRs.  In other words, the BPASS model that best fits
the level of photospheric line blanketing in the UV (which determines
the stellar metallicity) has an ionizing spectrum that yields
$\ha$-based SFRs consistent with those derived from the UV continuum
(see also discussion in \citealt{reddy23b}).  The fact that the model
can self-consistently explain the $\ha$ and non-ionizing UV
luminosities (or SFRs) implies that the predicted ionizing spectrum at
this metallicity must be reasonable.  Unless the physics relating
stellar opacity to the output ionizing spectrum is drastically
different at higher redshifts and/or at lower metallicities
($Z_\ast\la 0.002$), it stands to reason that the model predictions
for the shape and intensity of the ionizing spectrum in this regime
are likely to be reasonably accurate, an inference that appears to be
supported by indirect constraints on the ionizing spectra of
low-metallicity O stars (e.g., \citealt{telford23}).

The previous discussion centers on the ionizing spectrum between 1 and
4 Rydbergs (i.e., 13.6\,eV to 54.4\,eV), as it is photons with these
energies that dominate the ionization of hydrogen.  Several
investigations have suggested that even models that include the
effects of stellar binarity (such as the ones assumed here)---which
result in a harder and more intense ionizing spectrum at a fixed
metallicity \citep{eldridge17, stanway18}---predict an insufficient
number of He-ionizing photons ($>4$\,Rydbergs) to fully account for
the nebular $\heii$ emission seen in some high-redshift galaxies
(e.g., \citealt{shirazi12, senchyna17, schaerer19, nanayakkara19,
  stanway19}).  However, the models do reproduce the levels of nebular
$\heii$ emission in typical star-forming galaxies at $z\sim 2$
\citep{steidel16, reddy22}.  As with the discussion in the previous
sections, regardless of the metallicity dependence of the ionizing
spectrum at $>1$\,Rydberg, the large dispersion between $\log U$ and
metallicity (Figure~\ref{fig:uvsz}) points to factors other than
metallicity that drive the scatter in $U$ observed within the sample.

\subsubsection{Summary of Metallicity Considerations}
\label{sec:metallicitysummary}

We have considered the relationship between $\log U$ and $\log[Z_{\rm
    neb}/Z_\odot]$ inferred from photoionization modeling of the
$z=2.7-6.3$ galaxies in the sample.
There is a $1-1.5$\,dex variation in $\log U$ at a fixed $Z_{\rm neb}$
within our sample, a spread much larger than the measurement
uncertainties in $\log U$ for individual galaxies ($\la 0.2$\,dex).

Based on the predictions from the BPASS spectral synthesis models, the
expected variation in $Z_\ast$ within the sample is insufficient to
account for such a large variation in $U$.  We briefly discuss the
possibility that metallicity not only affects the hardness/intensity
of the ionizing spectrum, but also the gas (electron) density.
However, stellar feedback---which can result in higher pressures and
densities in $\hii$ regions---is expected to be {\em weaker} at lower
metallicities.  Moreover, dust attenuation of ionizing photons is not
expected to be significant for the typical (moderately-dusty) galaxies
analyzed in this work.  We also consider the possibility that the
hardness and/or intensity of the ionizing spectrum may increase more
rapidly with decreasing metallicity than the stellar population
synthesis model predictions indicate.  This possibility does not
appear to be supported by the constraints on the ionizing spectrum of
metal-poor $Z_\ast \simeq 0.001$ (i.e., $0.07\,Z_\odot$) typical
star-forming galaxies at $z\sim 2$, nor by the (albeit limited)
constraints on the ionizing spectrum of individual metal-poor O stars
\citep{telford23}.  Regardless of the effect of metallicity on either
the ionizing spectrum or gas density, the large scatter in $\log U$ at
a fixed $\log[Z_{\rm neb}/Z_\odot]$ suggests that factors other than
metallicity drive the variation in $\log U$ within the sample.

\begin{figure}
  \epsscale{1.1}
  \plotone{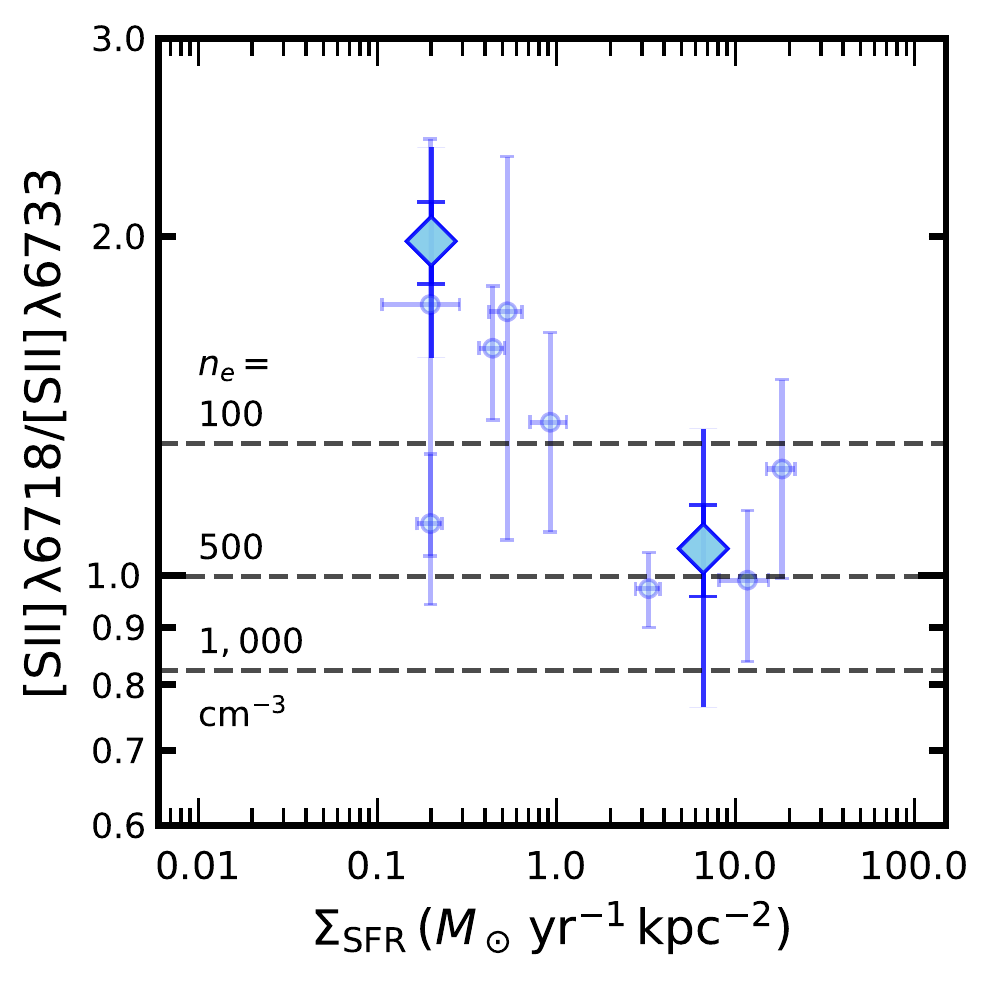}
    \caption{Variation of $\rsii$ with
      $\Sigma_{\rm SFR}$.  Individual galaxies with detections of
      $\sii$ are indicated by the small symbols.  The
      average $\langle \sii\rangle$ 
      measured from the composite spectra are shown by the large
      diamonds.  Capped
      vertical lines indicate measurement uncertainties, while
      uncapped vertical lines on the composite values include sample
      variance.}
    \label{fig:rsiisfrd}
\end{figure}

\begin{figure}
  \epsscale{1.1}
  \plotone{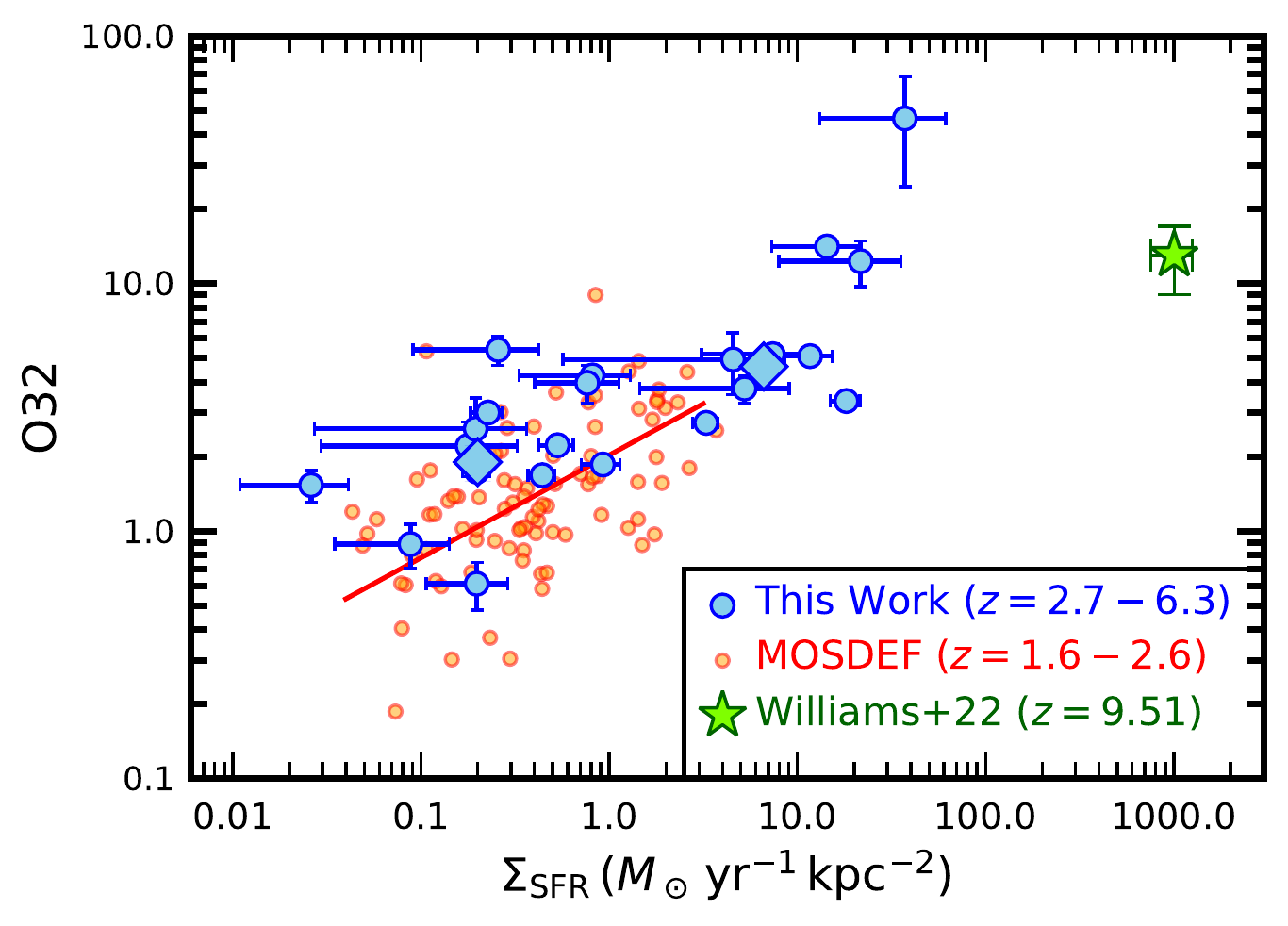}
    \caption{Variation of O32 with $\Sigma_{\rm SFR}$.  Individual
      galaxies with robust $\Sigma_{\rm SFR}$ measurements are
      indicated by the blue circles, and the large diamonds indicate
      the average values computed from composite spectra in two bins
      of O32.  Capped vertical lines indicate measurement
      uncertainties, while uncapped vertical lines on the composite
      values include sample variance.  In some cases, the error bars
      are smaller than the symbols.  Also shown are measurements from
      the MOSDEF survey (small red circles), along with the best-fit
      relation to those data (red line).  For comparison, the $z=9.51$
      lensed galaxy from \citet{williams22} is indicated by the green
      star.}
    \label{fig:oiisfrd}
\end{figure}

\begin{deluxetable*}{lccccc}
\tabletypesize{\footnotesize}
\tablewidth{0pc}
\tablecaption{$\Sigma_{\rm SFR}$ Subsamples}
\tablehead{
\colhead{$\Sigma_{\rm SFR}$ bin} &
\colhead{$N_{\rm comp}$\tablenotemark{a}} &
\colhead{$\langle z\rangle$\tablenotemark{b}} &
\colhead{$\langle \Sigma_{\rm SFR}\rangle$\tablenotemark{c}} &
\colhead{$\langle \rsii\rangle$\tablenotemark{d}} &
\colhead{$\langle {\rm O32}\rangle$\tablenotemark{e}}}
\startdata
low   & 11 & $3.570$ & $0.20\pm0.01$ & $1.981\pm0.165$ ($0.419$) & $1.899\pm 0.057$ ($0.314$)\\
high  & 11 & $4.876$ & $6.62\pm0.27$ & $1.057\pm0.099$ ($0.293$) & $4.620\pm 0.107$ ($0.455$) 
\enddata
\tablenotetext{a}{Number of galaxies with spectral coverage of $\rsii$ and $\Sigma_{\rm SFR}$ measurements, used to construct the composite spectra.}
\tablenotetext{b}{Mean redshift of galaxies in this bin.}
\tablenotetext{c}{Mean and uncertainty in mean $\Sigma_{\rm SFR}$ (in units of $M_\odot$\,yr${-1}$\,kpc$^{-2}$) for galaxies in this bin.}
\tablenotetext{d}{Mean and uncertainty in mean $\rsii$ for galaxies in this bin.  Numbers in parentheses include sample variance.}
\tablenotetext{e}{Mean and uncertainty in mean ${\rm O32}$ for galaxies in this bin.  Numbers in parentheses include sample variance.}
\label{tab:sfrd}
\end{deluxetable*}

\subsection{Star-Formation-Rate Surface Density}
\label{sec:sigmasfr}

The results of Section~\ref{sec:neq} imply a correlation between $n_e$
and $U$ (see also \citealt{shirazi14, bian16, papovich22, reddy23b}).
In turn, we can explore the factors responsible for variations in
$n_e$.  Previous efforts have established a connection between $n_e$
and gas density, or SFR surface density \citep{shirazi14, shimakawa15,
  bian16, jiang19, davies21, reddy23b}.  The {\em JWST} CEERS sample
allows us to examine this connection at significantly higher redshifts
($z\ga 2.7$) where galaxies may have a denser ISM on average.  For
this analysis, we considered the 22 galaxies in the sample that have
existing measurements of effective half-light radii ($R_{\rm eff}$)
from \citet{vanderwel14}.  The SFR surface density is then defined as:
\begin{equation}
\Sigma_{\rm SFR} = \frac{\sfrha}{2\pi R_{\rm eff}^2},
\end{equation}
and is expressed in units of $M_\odot$\,yr$^{-1}$\,kpc$^{-2}$.
Similar to the analysis presented in Section~\ref{sec:neq}, the
$\Sigma_{\rm SFR}$ subsample was divided into two equal-number bins of
$\Sigma_{\rm SFR}$, and $\langle \rsii\rangle$ and $\langle {\rm
  O32}\rangle$ were computed from the composite spectra constructed
for galaxies in the two bins.  The resulting values are listed in
Table~\ref{tab:sfrd} and shown in Figures~\ref{fig:rsiisfrd} and
\ref{fig:oiisfrd}.

The difference in $\langle\rsii\rangle$ between the low and high
$\Sigma_{\rm SFR}$ bins is significant at the $\approx 4\sigma$ level,
such that galaxies with higher $\Sigma_{\rm SFR}$ exhibit a lower
$\langle\rsii\rangle$ (or higher $\langle n_e \rangle \simeq
500$\,cm$^{-3}$), similar to the trend suggested by earlier studies.
As was the case for the relationship between $\rsii$ and O32
(Figure~\ref{fig:densityq}), an accounting of sample variance
indicates a large scatter in the relationship between $\rsii$ and
$\Sigma_{\rm SFR}$.

Of the correlations investigated in this analysis, the one between O32
and $\Sigma_{\rm SFR}$ appears to be the most significant
(Figure~\ref{fig:oiisfrd}): a Spearman correlation test indicates a
$p$ value of $p=1.97\times 10^{-5}$.  A similar and highly significant
correlation was also found by \citet{reddy23b} for $z\sim 2-4$
galaxies in the MOSDEF survey.  Data from that survey are also shown
in Figure~\ref{fig:oiisfrd}, along with the $z=9.51$ lensed galaxy
from \citet{williams22}, for comparison.  

\begin{figure}
  \epsscale{1.2}
  \plotone{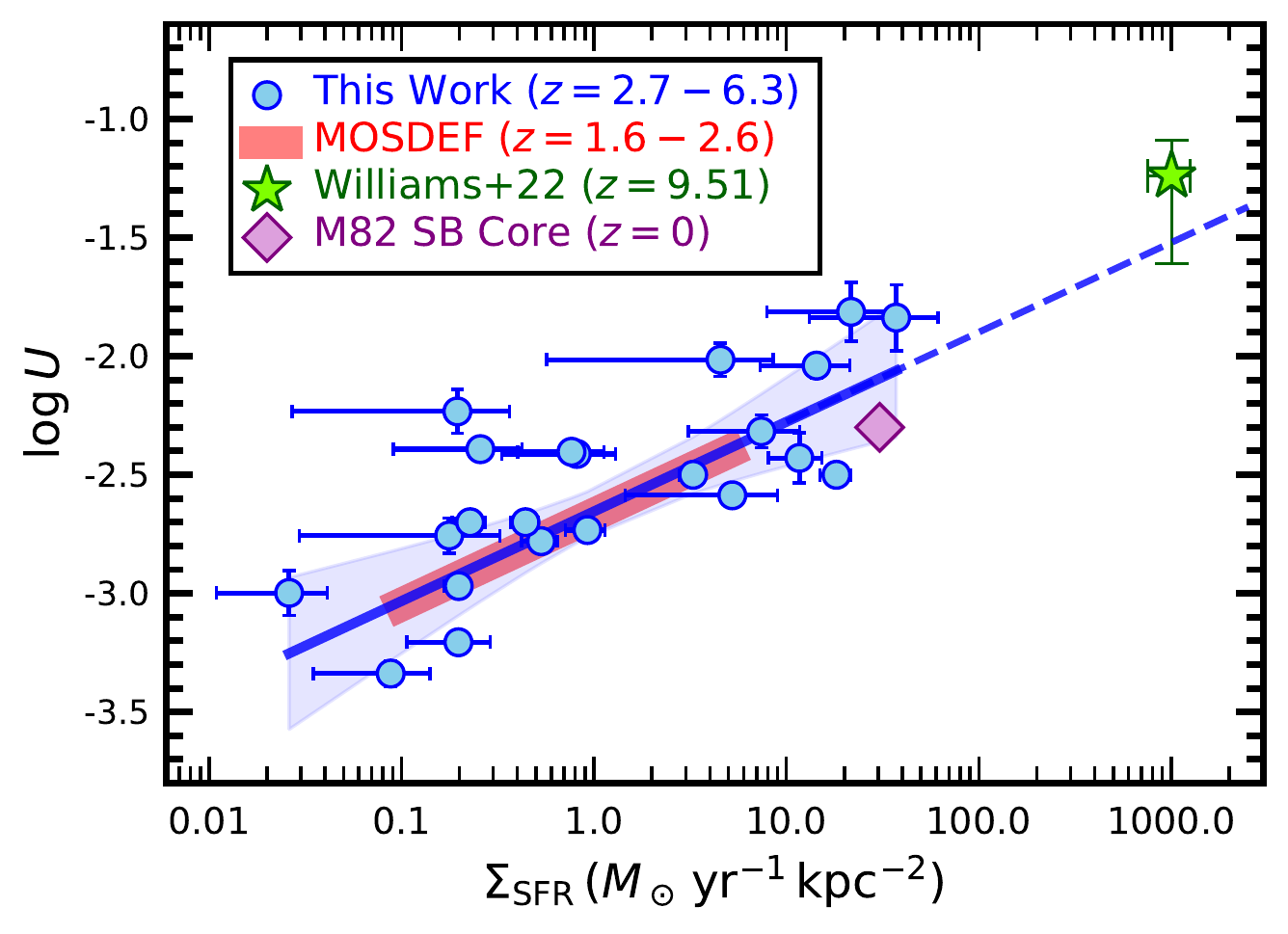}
    \caption{$\log U$ vs. $\Sigma_{\rm SFR}$ for the 22 galaxies in
      the sample with measurements of the latter.  A Spearman test
      indicates a probability $p=3.2\times 10^{-4}$ that the
      correlation occurs by random chance, implying a highly
      significant correlation.  A linear fit to the data indicates
      $\log U = (0.378\pm0.021)\log[\Sigma_{\rm SFR}/M_\odot\,{\rm
          yr}^{-1}\,{\rm kpc}^{-2}] + (-2.655\pm 0.011)$ (blue line),
      with the $2\sigma$ confidence interval shown by the blue shaded
      region.  The thick red line shows the best-fit correlation for
      $z=1.9-3.7$ galaxies in the MOSDEF survey \citep{reddy23b}, and
      is virtually identical to the relation for $z=2.7-6.3$ galaxies.
      The green star indicates the lensed $z=9.51$ galaxy from
      \citet{williams22}.  The purple diamond denotes the M82
      starburst core studied in \citet{forster01,forster03b}.}
    \label{fig:usfrd}
\end{figure}

Figure~\ref{fig:usfrd} directly shows the relationship between $\log
U$ inferred from photoionization modeling
(Section~\ref{sec:photoionization}) and $\Sigma_{\rm SFR}$ for the 22
galaxies with measurements of the latter.  Again, a Spearman
correlation test implies a highly significant correlation between
$\log U$ and $\Sigma_{\rm SFR}$, with $p=3.2\times 10^{-4}$.  A formal
linear fit to the sample of 22 galaxies yields the following relation
between $\log U$ and $\log[\Sigma_{\rm SFR}/M_\odot\,{\rm
    yr}^{-1}\,{\rm kpc}^{-2}]$:
\begin{eqnarray}
\log U & = & (0.378 \pm 0.021)\log\left[\frac{\Sigma_{\rm SFR}}{M_\odot\,{\rm yr}^{-1}\,{\rm kpc}^{-2}}\right] + \nonumber \\
& & (-2.655 \pm 0.011).
\label{eq:uvssigma}
\end{eqnarray}
A similar fit to lower-redshift ($z\sim 1.9-3.7$) galaxies from the
MOSDEF survey indicates a relation that is virtually identical to the
one found here, despite the differences in the way the two samples
were selected---i.e., rest-frame optical selection in the case of
MOSDEF versus a heterogeneous selection in the case of CEERS.  The
starburst core of M82 \citep{forster01,forster03b} is included in
Figure~\ref{fig:usfrd}, and is consistent with the trend between $U$
and $\Sigma_{\rm SFR}$ for the high-redshift samples.  For context,
the $z=9.51$ lensed galaxy in the RXJ2129 field reported by
\citet{williams22} has $U=-1.24$ and an extremely high $\Sigma_{\rm
  SFR} \simeq 1000$\,$M_\odot$\,yr$^{-1}$\,kpc$^{-2}$.  These values
place the $z \sim 9.5$ galaxy on the extrapolation of the $U$
vs. $\Sigma_{\rm SFR}$ relation established from the lower-redshift
samples.  This agreement, along with the virtually identical $U$
vs. $\Sigma_{\rm SFR}$ relations derived from the lower-redshift
MOSDEF sample ($\langle z\rangle \simeq 2.1$) and the current sample
($\langle z\rangle \simeq 4.4$), may indicate that the $U$
vs. $\Sigma_{\rm SFR}$ relation does not evolve with redshift.  If
additional measurements for high-$\Sigma_{\rm SFR}$ galaxies confirm
this lack of evolution, it would imply that the processes that couple
$U$ to $\Sigma_{\rm SFR}$ are independent of redshift and that
$\Sigma_{\rm SFR}$ is a key factor in modulating $U$ at any redshift.

\begin{figure}
  \epsscale{1.2}
  \plotone{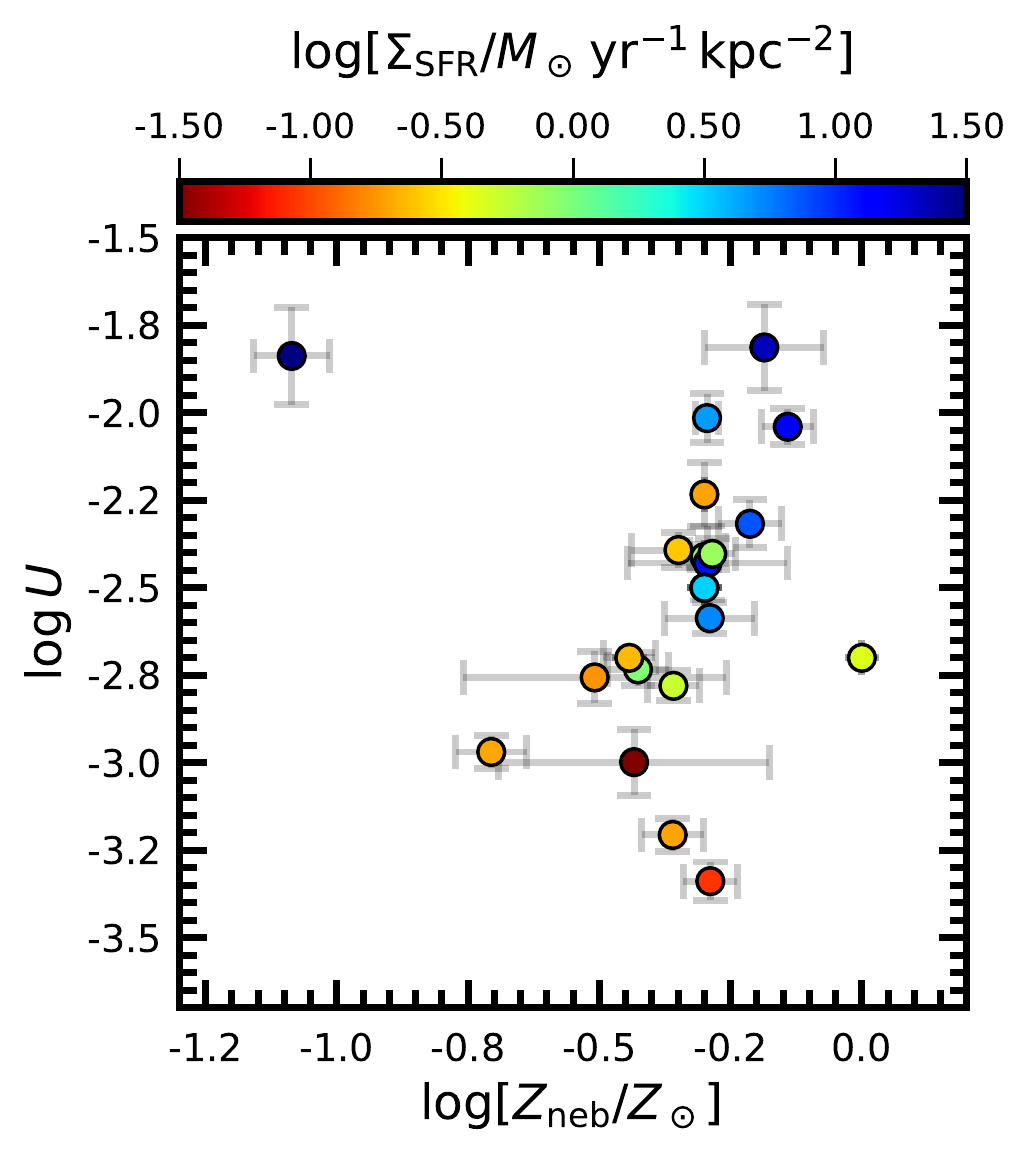}
    \caption{$\log U$ vs. $\log[Z_{\rm neb}/Z_\odot]$ for the 22
      galaxies in the sample with $\Sigma_{\rm SFR}$ measurements.
      The points are color coded by $\Sigma_{\rm SFR}$.}
    \label{fig:uvsz_colorcoded}
\end{figure}

Along these lines, we can examine whether $\Sigma_{\rm SFR}$ drives
the variation in $U$ found for galaxies in our sample.
Figure~\ref{fig:uvsz_colorcoded} displays the $\log U$ and
$\log[Z_{\rm neb}/Z_\odot]$ values for the 22 galaxies with
$\Sigma_{\rm SFR}$ measurements.  It is clear that the scatter in
$\log U$ at a given $\log[Z_{\rm neb}/Z_\odot]$ is related to
$\Sigma_{\rm SFR}$.  At a given $Z_{\rm neb}$, galaxies with the
highest $U$ also have high $\Sigma_{\rm SFR}$.

One result of our analysis is that $n_e$ appears to correlate
with---but displays a large scatter with respect to---O32.  
There is also substantial scatter between $Q$ and O32
(Figure~\ref{fig:densityq}).  On the other hand, we find highly
significant correlations between O32 and $\Sigma_{\rm SFR}$, and
between $\log U$ and $\Sigma_{\rm SFR}$.  This raises the possibility
that more than one of the parameters influencing $U$---namely $Q$,
$n_e$, $\epsilon$, and $f_{\rm esc}$---is connected to $\Sigma_{\rm
  SFR}$.  In other words, $\Sigma_{\rm SFR}$ may regulate more than
one of the aforementioned parameters affecting $U$, causing $U$ to
correlate tightly with $\Sigma_{\rm SFR}$.  We now explore this
possibility in the next few subsections.

\subsubsection{The Connection between $\Sigma_{\rm SFR}$ and $n_e$}

We have already discussed the difference in $\langle n_e\rangle$
inferred for low- and high-$\Sigma_{\rm SFR}$ galaxies.  A similar
trend of increasing $n_e$ with $\Sigma_{\rm SFR}$ has been established
in other low- and high-redshift samples, as noted above.  The
connection between $n_e$ and $\Sigma_{\rm SFR}$ may arise from the
densities of molecular clouds that, at least initially, dictate
densities in $\hii$ regions; the impact of stellar feedback on the
densities and pressures in $\hii$ regions; and/or pressure equilibrium
between $\hii$ regions and the ambient medium (e.g.,
\citealt{kaasinen17, davies21}).

\subsubsection{The Connection between $\Sigma_{\rm SFR}$ and $Q$}

In parallel, several investigations have highlighted the link between
SFR and $U$ \citep{nakajima14, kaasinen18, papovich22, reddy23b} and
its role in the redshift evolution of $U$ at a fixed stellar mass
(e.g., \citealt{reddy23b}).  Though we find no significant difference
in the $\langle Q\rangle$ for the low- and high-O32 subsamples
(Section~\ref{sec:neq} and Figure~\ref{fig:densityq}), an increase in
SFR would imply a corresponding increase in $Q$, and hence $U$.  The
dynamic range of SFR probed in the current sample is not sufficient to
readily observe an expected trend between SFR and $U$, highlighting
the need to conduct a similar analysis on samples covering a larger
dynamic range in SFR and other galaxy properties.  Regardless, since
the SFR is regulated by the gas supply and $Q \propto {\rm SFR}$, it
follows that $Q$ must also be related to gas (or SFR) surface density.

\subsubsection{The Connection between $\Sigma_{\rm SFR}$ and $\epsilon$}

Of the parameters that $U$ depends on, $\epsilon$ is the most
difficult to constrain at high redshift as it requires calculating the
volumes of line-emitting regions, which is beyond our capabilities
given the limited spatial resolution of the current observations.
Nevertheless, investigations of resolved local $\hii$ regions have
established a well-known inverse correlation between $n_e$ and $\hii$
region size \citep{kim01, hunt09}.  This ``density-size'' relation is
typically attributed to the preferential location of compact $\hii$
regions in denser molecular clouds \citep{larson81}, and the density
of such clouds is set by the gas or SFR surface density (e.g.,
\citealt{kennicutt21}).  Moreover, based on a sample of 58 $\hii$
regions in the disk of NGC 6946, \citet{cedres13} find a significant
anti-correlation between $\epsilon$ and $\hii$ region size.  If these
results are generalizable to other galaxies, they point to higher
volume filling fractions in compact $\hii$ regions forming in
dense-gas regions associated with high $\Sigma_{\rm SFR}$.  A
connection between $\epsilon$ and gas density is perhaps not
surprising if density inhomogeneities are more pronounced in
higher-density regions, leading to a larger filling fraction of dense
clumps within $\hii$ regions.  Consequently, $\epsilon$ may be
connected to $\Sigma_{\rm SFR}$.

\subsubsection{The Connection between $\Sigma_{\rm SFR}$ and $f_{\rm esc}$}
\label{sec:sfrdfesc}

Finally, we noted in Section~\ref{sec:densityneb} that an increase in
$f_{\rm esc}$ can lead to higher O32.  Based on direct measurements
locally (e.g., \citealt{gazagnes18}) and at $z\sim 2-3$
\citep{reddy16b, steidel18, reddy22}, $f_{\rm esc}$ is tightly
anti-correlated with the covering fraction of $\hi$.  The latter is
likely regulated by stellar feedback and supernovae explosions that
promote the formation of low column density or ionized channels in the
ISM (e.g., \citealt{gnedin08, ma16, kimm19, ma20, cen20, kakiichi21}).
The volumetric impact of such feedback is enhanced in regions of high
$\Sigma_{\rm SFR}$, particularly when coupled with a low gravitational
potential as is the case for low-stellar-mass galaxies at high
redshift \citep{reddy22}.  Thus, a connection between $f_{\rm esc}$
and $\Sigma_{\rm SFR}$ would follow.  

As discussed in Section~\ref{sec:densityneb}, as $f_{\rm esc}$
increases, the dependence of $U$ on $n_e$ is expected to become weaker
and eventually approach an inverse correlation.  If this is the case,
then significant outliers from the $U$ vs. $\Sigma_{\rm SFR}$ relation
(Figure~\ref{fig:usfrd}) may indicate objects where the positive
scaling between $U$ and $n_e$ breaks down.  This could occur if a
significant fraction of $4\pi$ steradians is subtended by
density-bounded regions (where $U$ would correlate negatively with
$n_e$), or if the overlapping $\hii$ regions in the galaxy result in a
more complicated scaling between $U$ and $n_e$.  Further data are
needed to test this potential avenue for identifying galaxies that may
be leaking a substantial fraction of ionizing photons or that may
otherwise have a complicated $\hii$ region geometry.

\subsubsection{Summary of $\Sigma_{\rm SFR}$ Considerations}

The previous subsections point to the possibility that $\Sigma_{\rm
  SFR}$ affects essentially all of the parameters ($n_e$, $Q$,
$\epsilon$, and $f_{\rm esc}$) that $U$ is sensitive to.  Based on the
previous discussion, all of these parameters may be expected to move
in the same direction as $\Sigma_{\rm SFR}$, as long as $U$ positively
correlates with $n_e$ (i.e., at low $f_{\rm esc}$).  For increasing
$f_{\rm esc}$---which would presumably indicate a high solid angle of
density-bounded sightlines---the dependence of $U$ on $n_e$ should
become noticeably weaker until an inverse correlation is established.
In that case, the increase in {\em apparent} $U$ will be dominated by
$f_{\rm esc}$ and $Q$.  The aforementioned behaviors provide a natural
explanation for why the scatter between any one of these parameters
and $U$ may be large, while their combination, which is sensitive to
$\Sigma_{\rm SFR}$, results in a highly significant correlation
between $U$ and $\Sigma_{\rm SFR}$.  This link between $U$ and
$\Sigma_{\rm SFR}$ explains much of the scatter in $U$ at a fixed
$Z_{\rm neb}$ (Figure~\ref{fig:uvsz_colorcoded})

\begin{figure}
  \epsscale{1.0}
  \plotone{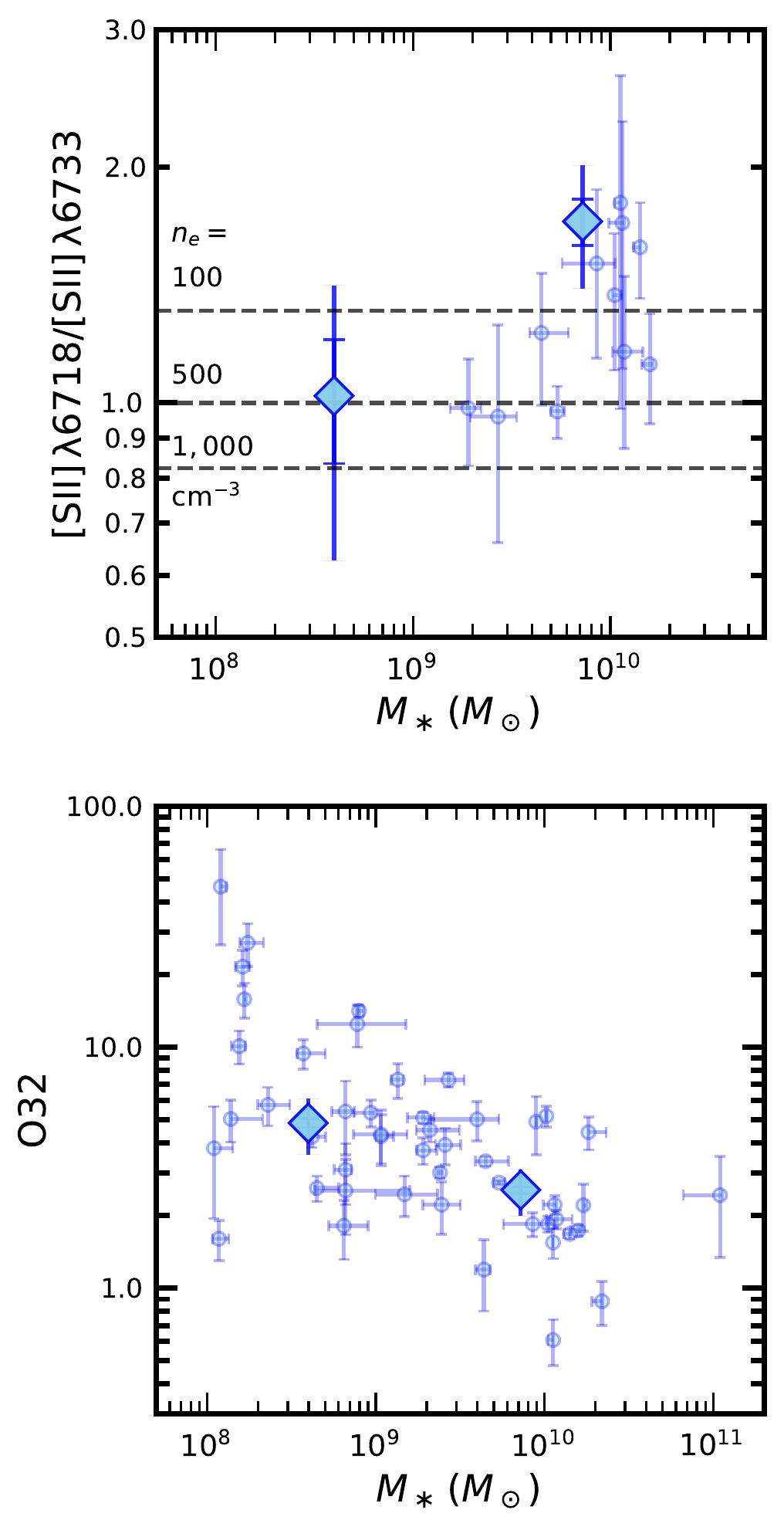}
    \caption{Variation of $\rsii$ (top) and ${\rm O32}$ (bottom) with
      $M_\ast$.  Individual galaxies with detections of $\sii$ and O32
      measurements are indicated by small circles in the top and
      bottom panels, respectively.  Mean measurements from composite
      spectra of galaxies are shown by the diamonds.  Measurement
      uncertainties for individual and composite values are indicated
      by the capped vertical lines.  Estimates of sample variance
      (including measurement error) for the composite values are
      indicated by the uncapped vertical lines.  In some cases, the
      error bars are smaller than the symbols.}
    \label{fig:mass}
\end{figure}

\begin{figure}
  \epsscale{1.2}
  \plotone{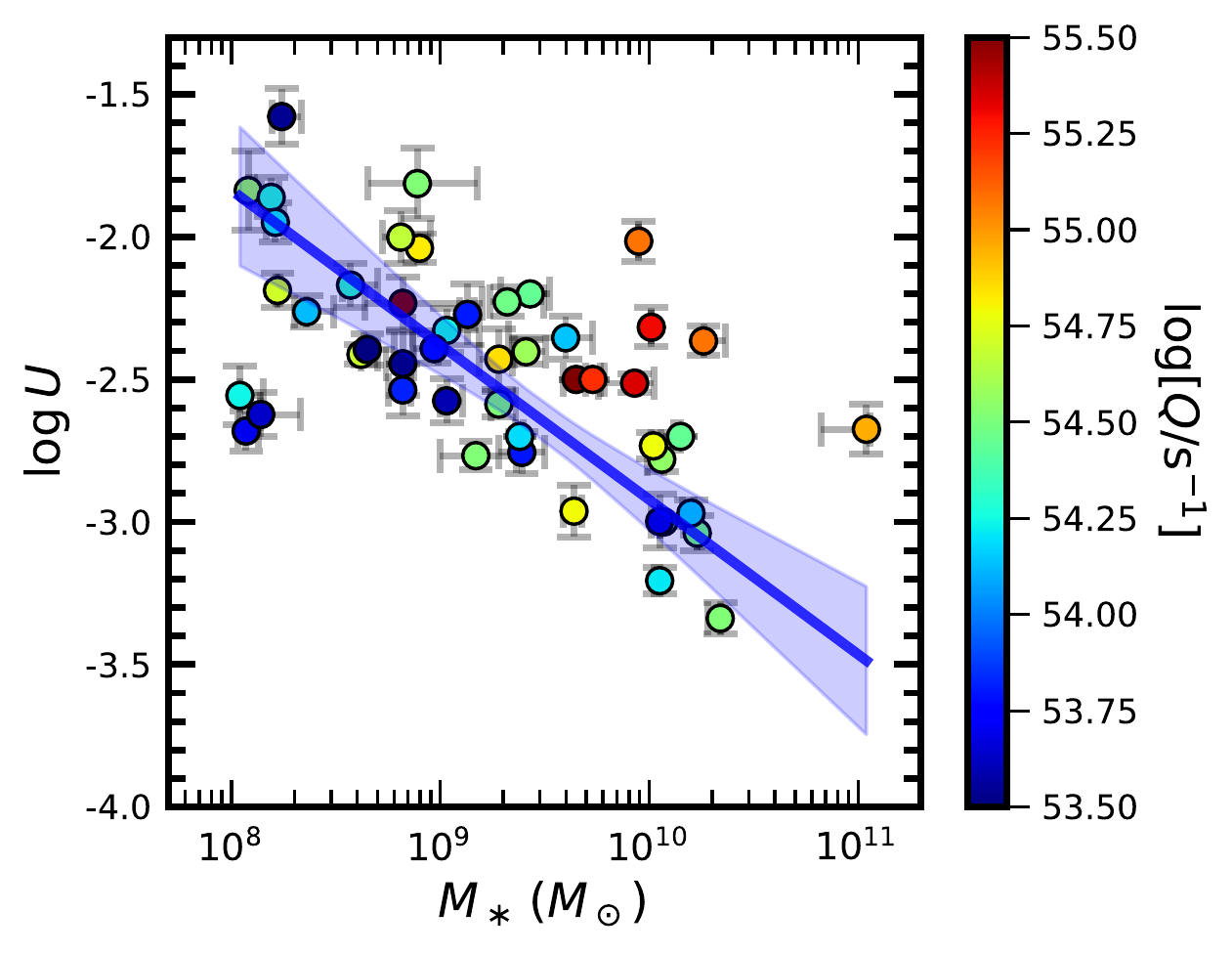}
    \caption{Relationship between $U$ and $M_\ast$.  Individual data
      points are color coded by $\log[Q/{\rm s}^{-1}]$.  The best-fit
      linear relation between $\log U$ and $\log[M_\ast/M_\odot]$, and
      its $2\sigma$ confidence interval, are indicated by the solid
      blue line and shaded blue region, respectively.  A formal fit
      gives $\log U = (-0.54+0.01)\log[M_\ast/M_\odot] +
      (2.51\pm0.14)$.}
    \label{fig:umass}
\end{figure}

\begin{deluxetable*}{lccccc}
\tabletypesize{\footnotesize}
\tablewidth{0pc}
\tablecaption{$M_\ast$ Subsamples}
\tablehead{
\colhead{$M_\ast$ bin} &
\colhead{$N_{\rm comp}$\tablenotemark{a}} &
\colhead{$\langle z\rangle$\tablenotemark{b}} &
\colhead{$\langle M_\ast\rangle$\tablenotemark{c}} &
\colhead{$\langle \rsii\rangle$\tablenotemark{d}} &
\colhead{$\langle {\rm O32}\rangle$\tablenotemark{e}}}
\startdata
low   & 24 & $4.766$ & $8.60\pm0.10$ & $1.019\pm0.184$ ($0.392$) & $4.847\pm 0.209$ ($1.270$)\\
high  & 24 & $4.011$ & $9.86\pm0.10$ & $1.704\pm0.116$ ($0.306$) & $2.554\pm 0.078$ ($0.561$) 
\enddata
\tablenotetext{a}{Number of galaxies with spectral coverage of $\rsii$, used to construct the composite spectra.}
\tablenotetext{b}{Mean redshift of galaxies in this bin.}
\tablenotetext{c}{Mean and uncertainty in mean $M_\ast$ (in units of $M_\odot$) for galaxies in this bin.}
\tablenotetext{d}{Mean and uncertainty in mean $\rsii$ for galaxies in this bin.  Numbers in parentheses include sample variance.}
\tablenotetext{e}{Mean and uncertainty in mean ${\rm O32}$ for galaxies in this bin.  Numbers in parentheses include sample variance.}
\label{tab:mass}
\end{deluxetable*}

\subsection{Stellar Mass Considerations}
\label{sec:mass}

Given the widespread use of stellar mass as a parameter in various
galaxy scaling relations, such as the SFR vs. $M_\ast$ and
mass-metallicity relations, it is useful to consider variations in $U$
and $n_e$ with $M_\ast$.  The sample of 48 galaxies was divided into
two bins of (low and high) $M_\ast$.  Composite spectra were
constructed for galaxies in each of the two bins, yielding the average
measurements of $\langle \rsii\rangle$ and $\langle {\rm O32}\rangle$
listed in Table~\ref{tab:mass} and displayed in Figure~\ref{fig:mass}.
The difference in $\langle\rsii\rangle$ measured between the two mass
bins is significant at the $3\sigma$ level, where galaxies with lower
stellar masses exhibit an average $\sii$ ratio indicative of higher
$\langle n_e\rangle \simeq 500$\,cm$^{-3}$.  This result contrasts
with those of \citet{shimakawa15} and \citet{sanders16b} for somewhat
lower-redshift galaxies ($z\sim 2.0-2.5$), where no significant
correlation was found between $n_e$ and $M_\ast$.  Along these lines,
accounting for sample variance suggests a large scatter in $\rsii$ at
a given $M_\ast$.  Indeed, \citet{gburek22} find an average $n_e$
consistent with the low-density limit ($\langle n_e\rangle \simeq
1$\,cm$^{-3}$) for a sample of lensed star-forming galaxies at $z\sim
2.3$ with a median stellar mass of $\log[M_\ast/M_\odot] \simeq 8.3$.
More precise estimates of $n_e$ for individual galaxies over a larger
dynamic range in $M_\ast$ will be needed to further probe the
relationship between $n_e$ and $M_\ast$.

Conversely, the relationship between ${\rm O32}$ and $M_\ast$ appears
to be more significant as inferred from both the individual and
composite measurements (see also \citealt{shapley23b}), consistent
with observations at lower redshift (e.g., \citealt{sanders16b}).  A
Spearman test on the individual measurements indicates a probability
$p=2.5\times 10^{-5}$ of a null correlation between O32 and $M_\ast$,
while the difference in $\langle {\rm O32}\rangle$ between the low-
and high-mass bins is significant at the $\simeq 10\sigma$ level.
Folding in the effect of sample variance suggests a fairly large
scatter in O32 at a given $M_\ast$, consistent with the spread of the
individual O32 measurements at a given $M_\ast$.

Figure~\ref{fig:umass} displays the relation between $\log U$ and
$M_\ast$ for the 48 galaxies in the sample.  Not surprisingly, as with
the relationship between O32 and $M_\ast$, a Spearman test indicates a
probability $p=2.0\times 10^{-5}$ that the observed relation is drawn
from an uncorrelated distribution of $\log U$ and $M_\ast$, implying a
highly significant ($\ga 4\sigma$) correlation.  For reference, the
best-fit linear correlation between $\log U$ and
$\log[M_\ast/M_\odot]$ is indicated in the figure, along with the
$2\sigma$ confidence interval on the fit.  Given the higher
$\Sigma_{\rm SFR}$ characteristic of low-mass galaxies (e.g.,
\citealt{shibuya15}), and the significant correlation between $\log U$
and $\Sigma_{\rm SFR}$ (see also \citealt{shimakawa15, reddy23b}), it
is not surprising that an anti-correlation between $\log U$ and
$M_\ast$ follows.  It is worth noting, however, that the lower SFRs
characteristic of low-mass galaxies implies a lower $Q$ (for a fixed
$\xi_{\rm ion}$) and hence lower $U$ if all other parameters are kept
fixed.  Variations in $Q$ are likely responsible for some of the
scatter seen in the relationship between $\log U$ and
$\log[M_\ast/M_\odot]$.  Specifically, Figure~\ref{fig:umass} shows
that galaxies with the highest $Q$ in the sample generally lie above
the mean relation between $\log U$ and $\log[M_\ast/M_\odot]$, while
those with lower $Q$ generally lie below this relation.  Thus the
scatter in the relation between $\log U$ and $\log[M_\ast/M_\odot]$ is
at least partly driven by variations in SFR.

Our analysis implies a strong anti-correlation between $\log U$ and
$\log[M_\ast/M_\odot]$ for galaxies at $z=2.7-6.3$, similar to trends
observed at lower redshifts.  The strength of this anti-correlation
and the apparently large scatter in the relationship between $n_e$ and
$M_\ast$ suggests that other factors that influence $U$, namely
$\epsilon$ and $f_{\rm esc}$, may also correlate inversely with
$M_\ast$.

\subsection{What Drives the Relationship between $U$ and $Z_{\rm neb}$ at High Redshift?}

We have argued that $\Sigma_{\rm SFR}$---which is tied to the gas
surface density via the Kennicutt-Schmidt relation
\citep{kennicutt98b, kennicutt21}---may play a central role in the
variation in $U$ seen at high redshift.  We should then address the
source of the anti-correlation between $U$ and $Z_{\rm neb}$, at least
when examined over a larger dynamic range in the latter.  We suggest
that this anti-correlation may stem from an increase in the average
$\Sigma_{\rm SFR}$ with decreasing $Z_{\rm neb}$.  While the data
shown in Figure~\ref{fig:uvsz_colorcoded} do not support this
hypothesis (i.e., there is a large spread in $\Sigma_{\rm SFR}$ at a
fixed $Z_{\rm neb}$), enlarging the current sample and/or including
other higher-redshift samples to probe the full range of $Z_{\rm neb}$
will be critical for determining how $Z_{\rm neb}$ depends on
$\Sigma_{\rm SFR}$.  The expectation of an inverse correlation between
$\Sigma_{\rm SFR}$ and $Z_{\rm neb}$ appears to be borne out by
spatially-resolved observations of local \citep{barrera18, baker23}
and high-redshift galaxies \citep{troncoso14}, and would naturally
follow from the anti-correlation between gas fraction and oxygen
abundance (e.g., \citealt{barrera18, sanders23a}).  It is also worth
noting that the lensed $z=9.51$ galaxy from \citet{williams22}, one of
the most metal-poor galaxies discovered to date with {\em JWST}, also
has an extremely high $\Sigma_{\rm SFR} \simeq
1000$\,$M_\odot$\,yr$^{-1}$\,kpc$^{-2}$.

The existence of an inverse correlation between $\Sigma_{\rm SFR}$ and
$Z_{\rm neb}$, and a positive correlation between $U$ and
$\Sigma_{\rm SFR}$ (Figure~\ref{fig:usfrd}), then imply an inverse
correlation between $U$ and $Z_{\rm neb}$.  In this case, the high
ionization parameters characteristic of galaxies with low oxygen
abundances is tied to the fact that such galaxies also have high
$\Sigma_{\rm SFR}$ on average (see also discussion in
\citealt{reddy23b}).  Future observations to constrain $U$, $Z_{\rm
  neb}$, and $\Sigma_{\rm SFR}$ or gas surface density in large samples
of high-redshift galaxies will elucidate the relative roles
of gas density and metallicity in shaping the distribution of $U$
found at high redshift.

\section{Conclusions}
\label{sec:conclusions}

We present the first statistical analysis of the connection between
$U$, $n_e$, and $\Sigma_{\rm SFR}$ for $z=2.7-6.3$ star-forming
galaxies.  The sample consists of 48 galaxies with CEERS {\em
  JWST}/NIRSpec spectroscopy of rest-frame optical emission lines, 22
of which have robustly measured sizes and hence $\Sigma_{\rm SFR}$
measurements.  Multiple Balmer emission lines were used to compute
dust-corrected line fluxes and ratios, which were then fit with
photoionization models to infer $U$ and $Z_{\rm neb}$.  Our analysis
indicates a correlation with a fairly large scatter between $n_e$ and
O32, suggesting that gas density may play a role in driving the
elevated $U$ observed for high-redshift galaxies.  The ionizing photon
rates, $Q$, appear to be uncorrelated with O32, likely owing to the
limited dynamic range in SFR probed by the sample.  

We discuss the role of metallicity in modulating $U$.  No significant
correlation is found between $U$ and $Z_{\rm neb}$ within the sample
(e.g., see also \citealt{topping20b}), but this is likely due to the
limited dynamic range in $Z_{\rm neb}$ probed by the current sample.
The expected range of stellar metallicities for galaxies in the sample
($Z_\ast \la 0.001$) implies that variations in the hardness or
intensity of the ionizing spectrum are unlikely to be the main driver
of $U$ in the sample.  We consider the possibility that metallicity
may affect $n_e$, and suggest that this is unlikely given the weakness
of stellar winds---and thus the reduced impact they would have on the
internal pressures and densities of $\hii$ regions---at low
metallicity.  We also consider the possibilities that dust diminishes
and softens the ionizing spectrum with increasing $Z_{\rm neb}$, or
that the stellar population synthesis models underpredict the increase
in hardness or intensity of the ionizing spectrum with decreasing
metallicity.  However, these possibilities are not supported by the
general agreement between ionization-rate-based and
non-ionizing-photon-based SFRs for lower redshift ($z\sim 2-3$)
galaxies, and the limited indirect constraints on the ionizing spectra
of individual low-metallicity O stars \citep{telford23}.

The data imply that $n_e$ may correlate with $\Sigma_{\rm SFR}$,
similar to findings from lower-redshift studies, though there is a
large scatter in this correlation.  On the other hand, we find a
relatively tight and highly significant correlation between $U$ and
$\Sigma_{\rm SFR}$, which appears to be redshift invariant at $z\sim
1.6-6.3$, and possibly up to $z\sim 9.5$.  We point to the possibility
that $\Sigma_{\rm SFR}$ may influence many (or all) of the factors
that $U$ is sensitive to: $n_e$, $Q$, the volume filling factor
$\epsilon$, and $f_{\rm esc}$.  If all of these factors move in tandem
with $\Sigma_{\rm SFR}$, as suggested by previous work, it would imply
a highly significant correlation between $U$ and $\Sigma_{\rm SFR}$,
consistent with our analysis.  Finally, we confirm the existence of an
anti-correlation between $U$ and $M_\ast$, similar to that established
in lower-redshift ($z\la 3$) galaxies.  This anti-correlation would
arise naturally from the higher $\Sigma_{\rm SFR}$ of low-mass
galaxies (e.g., \citealt{shibuya15}).  The main conclusion from this
study is that the variation in $U$ within the sample of $z=2.7-6.3$
galaxies is not due to metallicity, but rather driven by $\Sigma_{\rm
  SFR}$, or more fundamentally, the gas surface density.

We suggest a number of followup investigations to improve on the
existing analysis.  Larger and more representative samples of
high-redshift galaxies, covering larger dynamic ranges in properties
such as SFR and O32, will be needed to further test the strength of
the various correlations examined here.  Deeper observations with
higher $S/N$ spectra will enable the detection of weaker ionization-
and abundance-sensitive lines, yielding tighter constraints on $U$
and, in particular, $Z_{\rm neb}$ using line ratios independent of
those sensitive to $U$.  Likewise, higher spectral resolution will be
needed to resolve the $\oii$\,$\lambda\lambda 3727, 3730$ doublet,
which is arguably a more useful probe of $n_e$ given its strength and
slightly higher ionization potential relative to $\sii$.  Galaxy size
measurements for larger samples will be crucial for constraining
$\Sigma_{\rm SFR}$ and understanding the scatter between this property
and $U$.  Ultimately, direct measurements of gas surface densities
(e.g., with ALMA) will be needed to more directly examine the link
between gas density measured on kpc scales, and $n_e$ and $U$ which
are sensitive to the ionized gas on $\hii$-region scales.  More
extensive measurements of $n_e$ and $\epsilon$ in local $\hii$ regions
(and simulations of the dynamical evolution of such regions), and how
they relate to metallicity and SFR (or gas) surface density would
benefit the interpretation of the relevant correlations (or lack
thereof) seen at high redshift.  These improvements will allow us to
more rigorously quantify the role of metallicity and gas density in
explaining the elevated $U$ inferred for high-redshift galaxies.

\begin{acknowledgements}

We acknowledge the entire CEERS team for their effort to design and
execute this Early Release Science observational program.  We thank
Natascha F\"{o}rster Schreiber for useful conversations.  This work is
based on observations made with the NASA/ESA/CSA James Webb Space
Telescope.  The data were obtained from the Mikulski Archive for Space
Telescopes (MAST) at the Space Telescope Science Institute, which is
operated by the Association of Universities for Research in Astronomy,
Inc., under NASA contract NAS5-03127 for {\em JWST}.  The specific
observations analyzed can be accessed via \dataset[DOI
  10.17909/z7p0-8481]{https://archive.stsci.edu/doi/resolve/resolve.html?doi=10.17909/z7p0-8481}.
Support for this work was also provided through the NASA Hubble
Fellowship grant HST-HF2-51469.001-A awarded by the Space Telescope
Science Institute, which is operated by the Association of
Universities for Research in Astronomy, Incorporated, under NASA
contract NAS5-26555.  The Cosmic Dawn Center is funded by the Danish
National Research Foundation (DNRF) under grant \#140.  Cloud-based
data processing and file storage for this work is provided by the AWS
Cloud Credits for Research program.

\end{acknowledgements}



\begin{thebibliography}{}
\expandafter\ifx\csname natexlab\endcsname\relax\def\natexlab#1{#1}\fi

\bibitem[{{Ali}(2021)}]{ali21}
{Ali}, A.~A. 2021, \mnras, 501, 4136

\bibitem[{{Baker} {et~al.}(2023){Baker}, {Maiolino}, {Belfiore}, {Curti},
  {Bluck}, {Lin}, {Ellison}, {Thorp}, \& {Pan}}]{baker23}
{Baker}, W.~M., {Maiolino}, R., {Belfiore}, F., {et~al.} 2023, \mnras, 519,
  1149

\bibitem[{{Barrera-Ballesteros} {et~al.}(2018){Barrera-Ballesteros}, {Heckman},
  {S{\'a}nchez}, {Zakamska}, {Cleary}, {Zhu}, {Brinkmann}, {Drory}, \& {THE
  MaNGA TEAM}}]{barrera18}
{Barrera-Ballesteros}, J.~K., {Heckman}, T., {S{\'a}nchez}, S.~F., {et~al.}
  2018, \apj, 852, 74

\bibitem[{{Bian} {et~al.}(2016){Bian}, {Kewley}, {Dopita}, \&
  {Juneau}}]{bian16}
{Bian}, F., {Kewley}, L.~J., {Dopita}, M.~A., \& {Juneau}, S. 2016, \apj, 822,
  62

\bibitem[{{Bouwens} {et~al.}(2016){Bouwens}, {Smit}, {Labb{\'e}}, {Franx},
  {Caruana}, {Oesch}, {Stefanon}, \& {Rasappu}}]{bouwens16a}
{Bouwens}, R.~J., {Smit}, R., {Labb{\'e}}, I., {et~al.} 2016, \apj, 831, 176

\bibitem[{{Brinchmann} {et~al.}(2008){Brinchmann}, {Pettini}, \&
  {Charlot}}]{brinchmann08}
{Brinchmann}, J., {Pettini}, M., \& {Charlot}, S. 2008, \mnras, 385, 769

\bibitem[{{Brott} {et~al.}(2011){Brott}, {de Mink}, {Cantiello}, {Langer}, {de
  Koter}, {Evans}, {Hunter}, {Trundle}, \& {Vink}}]{brott11}
{Brott}, I., {de Mink}, S.~E., {Cantiello}, M., {et~al.} 2011, \aap, 530, A115

\bibitem[{{Buat} {et~al.}(2012){Buat}, {Noll}, {Burgarella}, {Giovannoli},
  {Charmandaris}, {Pannella}, {Hwang}, {Elbaz}, {Dickinson}, {Magdis}, {Reddy},
  \& {Murphy}}]{buat12}
{Buat}, V., {Noll}, S., {Burgarella}, D., {et~al.} 2012, \aap, 545, A141

\bibitem[{{Bunker} {et~al.}(2023){Bunker}, {Saxena}, {Cameron}, {Willott},
  {Curtis-Lake}, {Jakobsen}, {Carniani}, {Smit}, {Maiolino}, {Witstok},
  {Curti}, {D'Eugenio}, {Jones}, {Ferruit}, {Arribas}, {Charlot}, {Chevallard},
  {Giardino}, {de Graaff}, {Looser}, {Luetzgendorf}, {Maseda}, {Rawle}, {Rix},
  {Rodriguez Del Pino}, {Alberts}, {Egami}, {Eisenstein}, {Endsley},
  {Hainline}, {Hausen}, {Johnson}, {Rieke}, {Rieke}, {Robertson}, {Shivaei},
  {Stark}, {Sun}, {Tacchella}, {Tang}, {Williams}, {Willmer}, {Baker}, {Baum},
  {Bhatawdekar}, {Bowler}, {Boyett}, {Chen}, {Circosta}, {Helton}, {Ji}, {Lyu},
  {Nelson}, {Parlanti}, {Perna}, {Sandles}, {Scholtz}, {Suess}, {Topping},
  {Uebler}, {Wallace}, \& {Whitler}}]{bunker23}
{Bunker}, A.~J., {Saxena}, A., {Cameron}, A.~J., {et~al.} 2023, arXiv e-prints,
  arXiv:2302.07256

\bibitem[{{Calzetti} {et~al.}(2000){Calzetti}, {Armus}, {Bohlin}, {Kinney},
  {Koornneef}, \& {Storchi-Bergmann}}]{calzetti00}
{Calzetti}, D., {Armus}, L., {Bohlin}, R.~C., {et~al.} 2000, \apj, 533, 682

\bibitem[{{Cardelli} {et~al.}(1989){Cardelli}, {Clayton}, \&
  {Mathis}}]{cardelli89}
{Cardelli}, J.~A., {Clayton}, G.~C., \& {Mathis}, J.~S. 1989, \apj, 345, 245

\bibitem[{{Cedr{\'e}s} {et~al.}(2013){Cedr{\'e}s}, {Beckman}, {Bongiovanni},
  {Cepa}, {Asensio Ramos}, {Giammanco}, {Cabrera-Lavers}, \&
  {Alfaro}}]{cedres13}
{Cedr{\'e}s}, B., {Beckman}, J.~E., {Bongiovanni}, {\'A}., {et~al.} 2013,
  \apjl, 765, L24

\bibitem[{{Cen}(2020)}]{cen20}
{Cen}, R. 2020, \apjl, 889, L22

\bibitem[{{Chabrier}(2003)}]{chabrier03}
{Chabrier}, G. 2003, \pasp, 115, 763

\bibitem[{{Conroy} {et~al.}(2009){Conroy}, {Gunn}, \& {White}}]{conroy09}
{Conroy}, C., {Gunn}, J.~E., \& {White}, M. 2009, \apj, 699, 486

\bibitem[{{Cullen} {et~al.}(2019){Cullen}, {McLure}, {Dunlop}, {Khochfar},
  {Dav{\'e}}, {Amor{\'\i}n}, {Bolzonella}, {Carnall}, {Castellano}, {Cimatti},
  {Cirasuolo}, {Cresci}, {Fynbo}, {Fontanot}, {Gargiulo}, {Garilli}, {Guaita},
  {Hathi}, {Hibon}, {Mannucci}, {Marchi}, {McLeod}, {Pentericci}, {Pozzetti},
  {Shapley}, {Talia}, \& {Zamorani}}]{cullen19}
{Cullen}, F., {McLure}, R.~J., {Dunlop}, J.~S., {et~al.} 2019, \mnras, 487,
  2038

\bibitem[{{Cullen} {et~al.}(2021){Cullen}, {Shapley}, {McLure}, {Dunlop},
  {Sanders}, {Topping}, {Reddy}, {Amor{\'\i}n}, {Begley}, {Bolzonella},
  {Calabr{\`o}}, {Carnall}, {Castellano}, {Cimatti}, {Cirasuolo}, {Cresci},
  {Fontana}, {Fontanot}, {Garilli}, {Guaita}, {Hamadouche}, {Hathi},
  {Mannucci}, {McLeod}, {Pentericci}, {Saxena}, {Talia}, \&
  {Zamorani}}]{cullen21}
{Cullen}, F., {Shapley}, A.~E., {McLure}, R.~J., {et~al.} 2021, \mnras, 505,
  903

\bibitem[{{Daddi} {et~al.}(2007){Daddi}, {Dickinson}, {Morrison}, {Chary},
  {Cimatti}, {Elbaz}, {Frayer}, {Renzini}, {Pope}, {Alexander}, {Bauer},
  {Giavalisco}, {Huynh}, {Kurk}, \& {Mignoli}}]{daddi07a}
{Daddi}, E., {Dickinson}, M., {Morrison}, G., {et~al.} 2007, \apj, 670, 156

\bibitem[{{Davies} {et~al.}(2021){Davies}, {Schreiber}, {Genzel}, {Shimizu},
  {Davies}, {Schruba}, {Tacconi}, {{\"U}bler}, {Wisnioski}, {Wuyts}, {Fossati},
  {Herrera-Camus}, {Lutz}, {Mendel}, {Naab}, {Price}, {Renzini}, {Wilman},
  {Beifiori}, {Belli}, {Burkert}, {Chan}, {Contursi}, {Fabricius}, {Lee},
  {Saglia}, \& {Sternberg}}]{davies21}
{Davies}, R.~L., {Schreiber}, N.~M.~F., {Genzel}, R., {et~al.} 2021, \apj, 909,
  78

\bibitem[{{De Barros} {et~al.}(2016){De Barros}, {Reddy}, \&
  {Shivaei}}]{debarros16}
{De Barros}, S., {Reddy}, N., \& {Shivaei}, I. 2016, \apj, 820, 96

\bibitem[{{Dopita} \& {Evans}(1986)}]{dopita86}
{Dopita}, M.~A., \& {Evans}, I.~N. 1986, \apj, 307, 431

\bibitem[{{Dopita} {et~al.}(2006){Dopita}, {Fischera}, {Sutherland}, {Kewley},
  {Tuffs}, {Popescu}, {van Breugel}, {Groves}, \& {Leitherer}}]{dopita06}
{Dopita}, M.~A., {Fischera}, J., {Sutherland}, R.~S., {et~al.} 2006, \apj, 647,
  244

\bibitem[{{Draine}(2011)}]{draine11}
{Draine}, B.~T. 2011, \apj, 732, 100

\bibitem[{{Draine} {et~al.}(2007){Draine}, {Dale}, {Bendo}, {Gordon}, {Smith},
  {Armus}, {Engelbracht}, {Helou}, {Kennicutt}, {Li}, {Roussel}, {Walter},
  {Calzetti}, {Moustakas}, {Murphy}, {Rieke}, {Bot}, {Hollenbach}, {Sheth}, \&
  {Teplitz}}]{draine07b}
{Draine}, B.~T., {Dale}, D.~A., {Bendo}, G., {et~al.} 2007, \apj, 663, 866

\bibitem[{{Eldridge} {et~al.}(2017){Eldridge}, {Stanway}, {Xiao}, {McClelland},
  {Taylor}, {Ng}, {Greis}, \& {Bray}}]{eldridge17}
{Eldridge}, J.~J., {Stanway}, E.~R., {Xiao}, L., {et~al.} 2017, Publications of the Astronomical Society of Australia, 34, e058

\bibitem[{{Ferland} {et~al.}(2017){Ferland}, {Chatzikos}, {Guzm{\'a}n},
  {Lykins}, {van Hoof}, {Williams}, {Abel}, {Badnell}, {Keenan}, {Porter}, \&
  {Stancil}}]{ferland17}
{Ferland}, G.~J., {Chatzikos}, M., {Guzm{\'a}n}, F., {et~al.} 2017, Revista Mexicana de Astronomia y Astrofisica, 53,
  385

\bibitem[{{Finkelstein} {et~al.}(2022){Finkelstein}, {Bagley}, {Ferguson},
  {Wilkins}, {Kartaltepe}, {Papovich}, {Yung}, {Arrabal Haro}, {Behroozi},
  {Dickinson}, {Kocevski}, {Koekemoer}, {Larson}, {Le Bail}, {Morales},
  {Perez-Gonzalez}, {Burgarella}, {Dave}, {Hirschmann}, {Somerville}, {Wuyts},
  {Bromm}, {Casey}, {Fontana}, {Fujimoto}, {Gardner}, {Giavalisco}, {Grazian},
  {Grogin}, {Hathi}, {Hutchison}, {Jha}, {Jogee}, {Kewley}, {Kirkpatrick},
  {Long}, {Lotz}, {Pentericci}, {Pierel}, {Pirzkal}, {Ravindranath}, {Ryan},
  {Trump}, {Yang}, {Bhatawdekar}, {Bisigello}, {Buat}, {Calabro}, {Castellano},
  {Cleri}, {Cooper}, {Croton}, {Daddi}, {Dekel}, {Elbaz}, {Franco}, {Gawiser},
  {Holwerda}, {Huertas-Company}, {Jaskot}, {Leung}, {Lucas}, {Mobasher},
  {Pandya}, {Tacchella}, {Weiner}, \& {Zavala}}]{finkelstein22}
{Finkelstein}, S.~L., {Bagley}, M.~B., {Ferguson}, H.~C., {et~al.} 2022, arXiv
  e-prints, arXiv:2211.05792

\bibitem[{{F{\"o}rster Schreiber} {et~al.}(2001){F{\"o}rster Schreiber},
  {Genzel}, {Lutz}, {Kunze}, \& {Sternberg}}]{forster01}
{F{\"o}rster Schreiber}, N.~M., {Genzel}, R., {Lutz}, D., {Kunze}, D., \&
  {Sternberg}, A. 2001, \apj, 552, 544

\bibitem[{{F{\"o}rster Schreiber} {et~al.}(2003){F{\"o}rster Schreiber},
  {Genzel}, {Lutz}, \& {Sternberg}}]{forster03b}
{F{\"o}rster Schreiber}, N.~M., {Genzel}, R., {Lutz}, D., \& {Sternberg}, A.
  2003, \apj, 599, 193

\bibitem[{{Gazagnes} {et~al.}(2018){Gazagnes}, {Chisholm}, {Schaerer},
  {Verhamme}, {Rigby}, \& {Bayliss}}]{gazagnes18}
{Gazagnes}, S., {Chisholm}, J., {Schaerer}, D., {et~al.} 2018, \aap, 616, A29

\bibitem[{{Gburek} {et~al.}(2022){Gburek}, {Siana}, {Alavi}, {Emami},
  {Richard}, {Freeman}, {Stark}, \& {Snapp-Kolas}}]{gburek22}
{Gburek}, T., {Siana}, B., {Alavi}, A., {et~al.} 2022, arXiv e-prints,
  arXiv:2208.05976

\bibitem[{{Giammanco} {et~al.}(2005){Giammanco}, {Beckman}, \&
  {Cedr{\'e}s}}]{giammanco05}
{Giammanco}, C., {Beckman}, J.~E., \& {Cedr{\'e}s}, B. 2005, \aap, 438, 599

\bibitem[{{Gnedin} {et~al.}(2008){Gnedin}, {Kravtsov}, \& {Chen}}]{gnedin08}
{Gnedin}, N.~Y., {Kravtsov}, A.~V., \& {Chen}, H.-W. 2008, \apj, 672, 765

\bibitem[{{Gordon} {et~al.}(2003){Gordon}, {Clayton}, {Misselt}, {Landolt}, \&
  {Wolff}}]{gordon03}
{Gordon}, K.~D., {Clayton}, G.~C., {Misselt}, K.~A., {Landolt}, A.~U., \&
  {Wolff}, M.~J. 2003, \apj, 594, 279

\bibitem[{{Groves} {et~al.}(2008){Groves}, {Dopita}, {Sutherland}, {Kewley},
  {Fischera}, {Leitherer}, {Brandl}, \& {van Breugel}}]{groves08}
{Groves}, B., {Dopita}, M.~A., {Sutherland}, R.~S., {et~al.} 2008, \apjs, 176,
  438

\bibitem[{{Hunt} \& {Hirashita}(2009)}]{hunt09}
{Hunt}, L.~K., \& {Hirashita}, H. 2009, \aap, 507, 1327

\bibitem[{{Inoue}(2001)}]{inoue01a}
{Inoue}, A.~K. 2001, \aj, 122, 1788

\bibitem[{{Isobe} {et~al.}(2023){Isobe}, {Ouchi}, {Nakajima}, {Harikane},
  {Ono}, {Xu}, {Zhang}, \& {Umeda}}]{isobe23}
{Isobe}, Y., {Ouchi}, M., {Nakajima}, K., {et~al.} 2023, arXiv e-prints,
  arXiv:2301.06811

\bibitem[{{Jeong} {et~al.}(2020){Jeong}, {Shapley}, {Sanders}, {Runco},
  {Topping}, {Reddy}, {Kriek}, {Coil}, {Mobasher}, {Siana}, {Shivaei},
  {Freeman}, {Azadi}, {Price}, {Leung}, {Fetherolf}, {de Groot}, {Zick},
  {Fornasini}, \& {Barro}}]{jeong20}
{Jeong}, M.-S., {Shapley}, A.~E., {Sanders}, R.~L., {et~al.} 2020, \apjl, 902,
  L16

\bibitem[{{Jiang} {et~al.}(2019){Jiang}, {Malhotra}, {Yang}, \&
  {Rhoads}}]{jiang19}
{Jiang}, T., {Malhotra}, S., {Yang}, H., \& {Rhoads}, J.~E. 2019, \apj, 872,
  146

\bibitem[{{Kaasinen} {et~al.}(2017){Kaasinen}, {Bian}, {Groves}, {Kewley}, \&
  {Gupta}}]{kaasinen17}
{Kaasinen}, M., {Bian}, F., {Groves}, B., {Kewley}, L.~J., \& {Gupta}, A. 2017,
  \mnras, 465, 3220

\bibitem[{{Kaasinen} {et~al.}(2018){Kaasinen}, {Kewley}, {Bian}, {Groves},
  {Kashino}, {Silverman}, \& {Kartaltepe}}]{kaasinen18}
{Kaasinen}, M., {Kewley}, L., {Bian}, F., {et~al.} 2018, \mnras, 477, 5568

\bibitem[{{Kakiichi} \& {Gronke}(2021)}]{kakiichi21}
{Kakiichi}, K., \& {Gronke}, M. 2021, \apj, 908, 30

\bibitem[{{Kashino} \& {Inoue}(2019)}]{kashino19}
{Kashino}, D., \& {Inoue}, A.~K. 2019, \mnras, 486, 1053

\bibitem[{{Kennicutt}(1984)}]{kennicutt84}
{Kennicutt}, R.~C., J. 1984, \apj, 287, 116

\bibitem[{{Kennicutt} \& {De Los Reyes}(2021)}]{kennicutt21}
{Kennicutt}, Robert~C., J., \& {De Los Reyes}, M. A.~C. 2021, \apj, 908, 61

\bibitem[{{Kennicutt}(1998)}]{kennicutt98b}
{Kennicutt}, R.~C. 1998, \apj, 498, 541

\bibitem[{{Kim} \& {Koo}(2001)}]{kim01}
{Kim}, K.-T., \& {Koo}, B.-C. 2001, \apj, 549, 979

\bibitem[{{Kimm} {et~al.}(2019){Kimm}, {Blaizot}, {Garel}, {Michel-Dansac},
  {Katz}, {Rosdahl}, {Verhamme}, \& {Haehnelt}}]{kimm19}
{Kimm}, T., {Blaizot}, J., {Garel}, T., {et~al.} 2019, \mnras, 486, 2215

\bibitem[{{Kriek} {et~al.}(2009){Kriek}, {van Dokkum}, {Labb{\'e}}, {Franx},
  {Illingworth}, {Marchesini}, \& {Quadri}}]{kriek09}
{Kriek}, M., {van Dokkum}, P.~G., {Labb{\'e}}, I., {et~al.} 2009, \apj, 700,
  221

\bibitem[{{Krumholz} \& {Matzner}(2009)}]{krumholz09}
{Krumholz}, M.~R., \& {Matzner}, C.~D. 2009, \apj, 703, 1352

\bibitem[{{Kudritzki} \& {Puls}(2000)}]{kudritzki00}
{Kudritzki}, R.-P., \& {Puls}, J. 2000, \araa, 38, 613

\bibitem[{{Langer}(2012)}]{langer12}
{Langer}, N. 2012, \araa, 50, 107

\bibitem[{{Larson}(1981)}]{larson81}
{Larson}, R.~B. 1981, \mnras, 194, 809

\bibitem[{{Leitherer} {et~al.}(2014){Leitherer}, {Ekstr{\"o}m}, {Meynet},
  {Schaerer}, {Agienko}, \& {Levesque}}]{leitherer14}
{Leitherer}, C., {Ekstr{\"o}m}, S., {Meynet}, G., {et~al.} 2014, \apjs, 212, 14

\bibitem[{{Leitherer} \& {Heckman}(1995)}]{leitherer95}
{Leitherer}, C., \& {Heckman}, T.~M. 1995, \apjs, 96, 9

\bibitem[{{Levesque} \& {Richardson}(2014)}]{levesque14}
{Levesque}, E.~M., \& {Richardson}, M. L.~A. 2014, \apj, 780, 100

\bibitem[{{Liu} {et~al.}(2008){Liu}, {Shapley}, {Coil}, {Brinchmann}, \&
  {Ma}}]{liu08}
{Liu}, X., {Shapley}, A.~E., {Coil}, A.~L., {Brinchmann}, J., \& {Ma}, C.-P.
  2008, \apj, 678, 758

\bibitem[{{Ma} {et~al.}(2016){Ma}, {Hopkins}, {Kasen}, {Quataert},
  {Faucher-Gigu{\`e}re}, {Kere{\v{s}}}, {Murray}, \& {Strom}}]{ma16}
{Ma}, X., {Hopkins}, P.~F., {Kasen}, D., {et~al.} 2016, \mnras, 459, 3614

\bibitem[{{Ma} {et~al.}(2020){Ma}, {Quataert}, {Wetzel}, {Hopkins},
  {Faucher-Gigu{\`e}re}, \& {Kere{\v{s}}}}]{ma20}
{Ma}, X., {Quataert}, E., {Wetzel}, A., {et~al.} 2020, \mnras, 498, 2001

\bibitem[{{Masters} {et~al.}(2016){Masters}, {Faisst}, \& {Capak}}]{masters16}
{Masters}, D., {Faisst}, A., \& {Capak}, P. 2016, \apj, 828, 18

\bibitem[{{Nagao} {et~al.}(2006){Nagao}, {Maiolino}, \& {Marconi}}]{nagao06}
{Nagao}, T., {Maiolino}, R., \& {Marconi}, A. 2006, \aap, 459, 85

\bibitem[{{Naidu} {et~al.}(2022){Naidu}, {Matthee}, {Oesch}, {Conroy},
  {Sobral}, {Pezzulli}, {Hayes}, {Erb}, {Amor{\'\i}n}, {Gronke}, {Schaerer},
  {Tacchella}, {Kerutt}, {Paulino-Afonso}, {Calhau}, {Llerena}, \&
  {R{\"o}ttgering}}]{naidu22}
{Naidu}, R.~P., {Matthee}, J., {Oesch}, P.~A., {et~al.} 2022, \mnras, 510, 4582

\bibitem[{{Nakajima} \& {Ouchi}(2014)}]{nakajima14}
{Nakajima}, K., \& {Ouchi}, M. 2014, \mnras, 442, 900

\bibitem[{{Nakajima} {et~al.}(2013){Nakajima}, {Ouchi}, {Shimasaku},
  {Hashimoto}, {Ono}, \& {Lee}}]{nakajima13}
{Nakajima}, K., {Ouchi}, M., {Shimasaku}, K., {et~al.} 2013, \apj, 769, 3

\bibitem[{{Nanayakkara} {et~al.}(2019){Nanayakkara}, {Brinchmann}, {Boogaard},
  {Bouwens}, {Cantalupo}, {Feltre}, {Kollatschny}, {Marino}, {Maseda},
  {Matthee}, {Paalvast}, {Richard}, \& {Verhamme}}]{nanayakkara19}
{Nanayakkara}, T., {Brinchmann}, J., {Boogaard}, L., {et~al.} 2019, \aap, 624,
  A89

\bibitem[{{Osterbrock} \& {Flather}(1959)}]{osterbrock59}
{Osterbrock}, D., \& {Flather}, E. 1959, \apj, 129, 26

\bibitem[{{Pacifici} {et~al.}(2015){Pacifici}, {da Cunha}, {Charlot}, {Rix},
  {Fumagalli}, {Wel}, {Franx}, {Maseda}, {van Dokkum}, {Brammer}, {Momcheva},
  {Skelton}, {Whitaker}, {Leja}, {Lundgren}, {Kassin}, \& {Yi}}]{pacifici15}
{Pacifici}, C., {da Cunha}, E., {Charlot}, S., {et~al.} 2015, \mnras, 447, 786

\bibitem[{{Pannella} {et~al.}(2015){Pannella}, {Elbaz}, {Daddi}, {Dickinson},
  {Hwang}, {Schreiber}, {Strazzullo}, {Aussel}, {Bethermin}, {Buat},
  {Charmandaris}, {Cibinel}, {Juneau}, {Ivison}, {Le Borgne}, {Le Floc'h},
  {Leiton}, {Lin}, {Magdis}, {Morrison}, {Mullaney}, {Onodera}, {Renzini},
  {Salim}, {Sargent}, {Scott}, {Shu}, \& {Wang}}]{pannella15}
{Pannella}, M., {Elbaz}, D., {Daddi}, E., {et~al.} 2015, \apj, 807, 141

\bibitem[{{Papovich} {et~al.}(2011){Papovich}, {Finkelstein}, {Ferguson},
  {Lotz}, \& {Giavalisco}}]{papovich11}
{Papovich}, C., {Finkelstein}, S.~L., {Ferguson}, H.~C., {Lotz}, J.~M., \&
  {Giavalisco}, M. 2011, \mnras, 412, 1123

\bibitem[{{Papovich} {et~al.}(2022){Papovich}, {Simons}, {Estrada-Carpenter},
  {Matharu}, {Momcheva}, {Trump}, {Backhaus}, {Brammer}, {Cleri},
  {Finkelstein}, {Giavalisco}, {Ji}, {Jung}, {Kewley}, {Nicholls}, {Pirzkal},
  {Rafelski}, \& {Weiner}}]{papovich22}
{Papovich}, C., {Simons}, R.~C., {Estrada-Carpenter}, V., {et~al.} 2022, \apj,
  937, 22

\bibitem[{{P{\'e}rez-Montero}(2014)}]{perez14}
{P{\'e}rez-Montero}, E. 2014, \mnras, 441, 2663

\bibitem[{{P{\'e}rez-Montero} {et~al.}(2007){P{\'e}rez-Montero}, {H{\"a}gele},
  {Contini}, \& {D{\'\i}az}}]{perez07}
{P{\'e}rez-Montero}, E., {H{\"a}gele}, G.~F., {Contini}, T., \& {D{\'\i}az},
  {\'A}.~I. 2007, \mnras, 381, 125

\bibitem[{{Reddy} {et~al.}(2010){Reddy}, {Erb}, {Pettini}, {Steidel}, \&
  {Shapley}}]{reddy10}
{Reddy}, N.~A., {Erb}, D.~K., {Pettini}, M., {Steidel}, C.~C., \& {Shapley},
  A.~E. 2010, \apj, 712, 1070

\bibitem[{{Reddy} {et~al.}(2012){Reddy}, {Pettini}, {Steidel}, {Shapley},
  {Erb}, \& {Law}}]{reddy12b}
{Reddy}, N.~A., {Pettini}, M., {Steidel}, C.~C., {et~al.} 2012, \apj, 754, 25

\bibitem[{{Reddy} {et~al.}(2006){Reddy}, {Steidel}, {Fadda}, {Yan}, {Pettini},
  {Shapley}, {Erb}, \& {Adelberger}}]{reddy06a}
{Reddy}, N.~A., {Steidel}, C.~C., {Fadda}, D., {et~al.} 2006, \apj, 644, 792

\bibitem[{{Reddy} {et~al.}(2016){Reddy}, {Steidel}, {Pettini},
  {Bogosavljevi{\'c}}, \& {Shapley}}]{reddy16b}
{Reddy}, N.~A., {Steidel}, C.~C., {Pettini}, M., {Bogosavljevi{\'c}}, M., \&
  {Shapley}, A.~E. 2016, \apj, 828, 108

\bibitem[{{Reddy} {et~al.}(2023{\natexlab{a}}){Reddy}, {Topping}, {Sanders},
  {Shapley}, \& {Brammer}}]{reddy23a}
{Reddy}, N.~A., {Topping}, M.~W., {Sanders}, R.~L., {Shapley}, A.~E., \&
  {Brammer}, G. 2023{\natexlab{a}}, arXiv e-prints, arXiv:2301.07249

\bibitem[{{Reddy} {et~al.}(2018){Reddy}, {Oesch}, {Bouwens}, {Montes},
  {Illingworth}, {Steidel}, {van Dokkum}, {Atek}, {Carollo}, {Cibinel},
  {Holden}, {Labb{\'e}}, {Magee}, {Morselli}, {Nelson}, \&
  {Wilkins}}]{reddy18a}
{Reddy}, N.~A., {Oesch}, P.~A., {Bouwens}, R.~J., {et~al.} 2018, \apj, 853, 56

\bibitem[{{Reddy} {et~al.}(2020){Reddy}, {Shapley}, {Kriek}, {Steidel},
  {Shivaei}, {Sanders}, {Mobasher}, {Coil}, {Siana}, {Freeman}, {Azadi},
  {Fetherolf}, {Leung}, {Price}, \& {Zick}}]{reddy20}
{Reddy}, N.~A., {Shapley}, A.~E., {Kriek}, M., {et~al.} 2020, \apj, 902, 123

\bibitem[{{Reddy} {et~al.}(2022){Reddy}, {Topping}, {Shapley}, {Steidel},
  {Sanders}, {Du}, {Coil}, {Mobasher}, {Price}, \& {Shivaei}}]{reddy22}
{Reddy}, N.~A., {Topping}, M.~W., {Shapley}, A.~E., {et~al.} 2022, \apj, 926,
  31

\bibitem[{{Reddy} {et~al.}(2023{\natexlab{b}}){Reddy}, {Sanders}, {Shapley},
  {Topping}, {Kriek}, {Coil}, {Mobasher}, {Siana}, \& {Rezaee}}]{reddy23b}
{Reddy}, N.~A., {Sanders}, R.~L., {Shapley}, A.~E., {et~al.}
  2023{\natexlab{b}}, arXiv e-prints, arXiv:2302.10213

\bibitem[{{R{\'e}my-Ruyer} {et~al.}(2014){R{\'e}my-Ruyer}, {Madden},
  {Galliano}, {Galametz}, {Takeuchi}, {Asano}, {Zhukovska}, {Lebouteiller},
  {Cormier}, {Jones}, {Bocchio}, {Baes}, {Bendo}, {Boquien}, {Boselli},
  {DeLooze}, {Doublier-Pritchard}, {Hughes}, {Karczewski}, \&
  {Spinoglio}}]{ruyer14}
{R{\'e}my-Ruyer}, A., {Madden}, S.~C., {Galliano}, F., {et~al.} 2014, \aap,
  563, A31

\bibitem[{{Robertson} {et~al.}(2013){Robertson}, {Furlanetto}, {Schneider},
  {Charlot}, {Ellis}, {Stark}, {McLure}, {Dunlop}, {Koekemoer}, {Schenker},
  {Ouchi}, {Ono}, {Curtis-Lake}, {Rogers}, {Bowler}, \&
  {Cirasuolo}}]{robertson13}
{Robertson}, B.~E., {Furlanetto}, S.~R., {Schneider}, E., {et~al.} 2013, \apj,
  768, 71

\bibitem[{{Runco} {et~al.}(2021){Runco}, {Shapley}, {Sanders}, {Topping},
  {Kriek}, {Reddy}, {Coil}, {Mobasher}, {Siana}, {Freeman}, {Shivaei}, {Azadi},
  {Price}, {Leung}, {Fetherolf}, {de Groot}, {Zick}, {Fornasini}, \&
  {Barro}}]{runco21}
{Runco}, J.~N., {Shapley}, A.~E., {Sanders}, R.~L., {et~al.} 2021, \mnras, 502,
  2600

\bibitem[{{Sanders} {et~al.}(2023{\natexlab{a}}){Sanders}, {Shapley},
  {Topping}, {Reddy}, \& {Brammer}}]{sanders23c}
{Sanders}, R.~L., {Shapley}, A.~E., {Topping}, M.~W., {Reddy}, N.~A., \&
  {Brammer}, G.~B. 2023{\natexlab{a}}, arXiv e-prints, arXiv:2301.06696

\bibitem[{{Sanders} {et~al.}(2016{\natexlab{a}}){Sanders}, {Shapley}, {Kriek},
  {Reddy}, {Freeman}, {Coil}, {Siana}, {Mobasher}, {Shivaei}, {Price}, \& {de
  Groot}}]{sanders16b}
{Sanders}, R.~L., {Shapley}, A.~E., {Kriek}, M., {et~al.} 2016{\natexlab{a}},
  \apjl, 825, L23

\bibitem[{{Sanders} {et~al.}(2016{\natexlab{b}}){Sanders}, {Shapley}, {Kriek},
  {Reddy}, {Freeman}, {Coil}, {Siana}, {Mobasher}, {Shivaei}, {Price}, \& {de
  Groot}}]{sanders16a}
---. 2016{\natexlab{b}}, \apj, 816, 23

\bibitem[{{Sanders} {et~al.}(2020){Sanders}, {Shapley}, {Reddy}, {Kriek},
  {Siana}, {Coil}, {Mobasher}, {Shivaei}, {Freeman}, {Azadi}, {Price}, {Leung},
  {Fetherolf}, {de Groot}, {Zick}, {Fornasini}, \& {Barro}}]{sanders20}
{Sanders}, R.~L., {Shapley}, A.~E., {Reddy}, N.~A., {et~al.} 2020, \mnras, 491,
  1427

\bibitem[{{Sanders} {et~al.}(2023{\natexlab{b}}){Sanders}, {Shapley}, {Jones},
  {Shivaei}, {Popping}, {Reddy}, {Dav{\'e}}, {Price}, {Mobasher}, {Kriek},
  {Coil}, \& {Siana}}]{sanders23a}
{Sanders}, R.~L., {Shapley}, A.~E., {Jones}, T., {et~al.} 2023{\natexlab{b}},
  \apj, 942, 24

\bibitem[{{Schaerer} {et~al.}(2019){Schaerer}, {Fragos}, \&
  {Izotov}}]{schaerer19}
{Schaerer}, D., {Fragos}, T., \& {Izotov}, Y.~I. 2019, \aap, 622, L10

\bibitem[{{Senchyna} {et~al.}(2017){Senchyna}, {Stark}, {Vidal-Garc{\'\i}a},
  {Chevallard}, {Charlot}, {Mainali}, {Jones}, {Wofford}, {Feltre}, \&
  {Gutkin}}]{senchyna17}
{Senchyna}, P., {Stark}, D.~P., {Vidal-Garc{\'\i}a}, A., {et~al.} 2017, \mnras,
  472, 2608

\bibitem[{{Shapley} {et~al.}(2023{\natexlab{a}}){Shapley}, {Reddy}, {Sanders},
  {Topping}, \& {Brammer}}]{shapley23b}
{Shapley}, A.~E., {Reddy}, N.~A., {Sanders}, R.~L., {Topping}, M.~W., \&
  {Brammer}, G.~B. 2023{\natexlab{a}}, arXiv e-prints, arXiv:2303.00410

\bibitem[{{Shapley} {et~al.}(2023{\natexlab{b}}){Shapley}, {Sanders}, {Reddy},
  {Topping}, \& {Brammer}}]{shapley23a}
{Shapley}, A.~E., {Sanders}, R.~L., {Reddy}, N.~A., {Topping}, M.~W., \&
  {Brammer}, G.~B. 2023{\natexlab{b}}, arXiv e-prints, arXiv:2301.03241

\bibitem[{{Shibuya} {et~al.}(2015){Shibuya}, {Ouchi}, \&
  {Harikane}}]{shibuya15}
{Shibuya}, T., {Ouchi}, M., \& {Harikane}, Y. 2015, \apjs, 219, 15

\bibitem[{{Shimakawa} {et~al.}(2015){Shimakawa}, {Kodama}, {Steidel}, {Tadaki},
  {Tanaka}, {Strom}, {Hayashi}, {Koyama}, {Suzuki}, \&
  {Yamamoto}}]{shimakawa15}
{Shimakawa}, R., {Kodama}, T., {Steidel}, C.~C., {et~al.} 2015, \mnras, 451,
  1284

\bibitem[{{Shirazi} \& {Brinchmann}(2012)}]{shirazi12}
{Shirazi}, M., \& {Brinchmann}, J. 2012, \mnras, 421, 1043

\bibitem[{{Shirazi} {et~al.}(2014){Shirazi}, {Brinchmann}, \&
  {Rahmati}}]{shirazi14}
{Shirazi}, M., {Brinchmann}, J., \& {Rahmati}, A. 2014, \apj, 787, 120

\bibitem[{{Shivaei} {et~al.}(2015){Shivaei}, {Reddy}, {Steidel}, \&
  {Shapley}}]{shivaei15a}
{Shivaei}, I., {Reddy}, N.~A., {Steidel}, C.~C., \& {Shapley}, A.~E. 2015,
  \apj, 804, 149

\bibitem[{{Shivaei} {et~al.}(2016){Shivaei}, {Kriek}, {Reddy}, {Shapley},
  {Barro}, {Conroy}, {Coil}, {Freeman}, {Mobasher}, {Siana}, {Sanders},
  {Price}, {Azadi}, {Pasha}, \& {Inami}}]{shivaei16}
{Shivaei}, I., {Kriek}, M., {Reddy}, N.~A., {et~al.} 2016, \apjl, 820, L23

\bibitem[{{Shivaei} {et~al.}(2018){Shivaei}, {Reddy}, {Siana}, {Shapley},
  {Kriek}, {Mobasher}, {Freeman}, {Sanders}, {Coil}, {Price}, {Fetherolf},
  {Azadi}, {Leung}, \& {Zick}}]{shivaei18}
{Shivaei}, I., {Reddy}, N.~A., {Siana}, B., {et~al.} 2018, \apj, 855, 42

\bibitem[{{Shivaei} {et~al.}(2020){Shivaei}, {Reddy}, {Rieke}, {Shapley},
  {Kriek}, {Battisti}, {Mobasher}, {Sanders}, {Fetherolf}, {Azadi}, {Coil},
  {Freeman}, {de Groot}, {Leung}, {Price}, {Siana}, \& {Zick}}]{shivaei20a}
{Shivaei}, I., {Reddy}, N., {Rieke}, G., {et~al.} 2020, \apj, 899, 117

\bibitem[{{Stanway} \& {Eldridge}(2018)}]{stanway18}
{Stanway}, E.~R., \& {Eldridge}, J.~J. 2018, \mnras, 479, 75

\bibitem[{{Stanway} \& {Eldridge}(2019)}]{stanway19}
---. 2019, \aap, 621, A105

\bibitem[{{Steidel} {et~al.}(2018){Steidel}, {Bogosavljevi{\'c}}, {Shapley},
  {Reddy}, {Rudie}, {Pettini}, {Trainor}, \& {Strom}}]{steidel18}
{Steidel}, C.~C., {Bogosavljevi{\'c}}, M., {Shapley}, A.~E., {et~al.} 2018,
  \apj, 869, 123

\bibitem[{{Steidel} {et~al.}(2016){Steidel}, {Strom}, {Pettini}, {Rudie},
  {Reddy}, \& {Trainor}}]{steidel16}
{Steidel}, C.~C., {Strom}, A.~L., {Pettini}, M., {et~al.} 2016, \apj, 826, 159

\bibitem[{{Strom} {et~al.}(2018){Strom}, {Steidel}, {Rudie}, {Trainor}, \&
  {Pettini}}]{strom18}
{Strom}, A.~L., {Steidel}, C.~C., {Rudie}, G.~C., {Trainor}, R.~F., \&
  {Pettini}, M. 2018, \apj, 868, 117

\bibitem[{{Strom} {et~al.}(2017){Strom}, {Steidel}, {Rudie}, {Trainor},
  {Pettini}, \& {Reddy}}]{strom17}
{Strom}, A.~L., {Steidel}, C.~C., {Rudie}, G.~C., {et~al.} 2017, \apj, 836, 164

\bibitem[{{Tang} {et~al.}(2023){Tang}, {Stark}, {Chen}, {Mason}, {Topping},
  {Endsley}, {Senchyna}, {Plat}, {Lu}, {Whitler}, {Robertson}, \&
  {Charlot}}]{tang23}
{Tang}, M., {Stark}, D.~P., {Chen}, Z., {et~al.} 2023, arXiv e-prints,
  arXiv:2301.07072

\bibitem[{{Telford} {et~al.}(2023){Telford}, {McQuinn}, {Chisholm}, \&
  {Berg}}]{telford23}
{Telford}, O.~G., {McQuinn}, K. B.~W., {Chisholm}, J., \& {Berg}, D.~A. 2023,
  \apj, 943, 65

\bibitem[{{Theios} {et~al.}(2019){Theios}, {Steidel}, {Strom}, {Rudie},
  {Trainor}, \& {Reddy}}]{theios19}
{Theios}, R.~L., {Steidel}, C.~C., {Strom}, A.~L., {et~al.} 2019, \apj, 871,
  128

\bibitem[{{Topping} {et~al.}(2020{\natexlab{a}}){Topping}, {Shapley}, {Reddy},
  {Sanders}, {Coil}, {Kriek}, {Mobasher}, \& {Siana}}]{topping20b}
{Topping}, M.~W., {Shapley}, A.~E., {Reddy}, N.~A., {et~al.}
  2020{\natexlab{a}}, \mnras, 499, 1652

\bibitem[{{Topping} {et~al.}(2020{\natexlab{b}}){Topping}, {Shapley}, {Reddy},
  {Sanders}, {Coil}, {Kriek}, {Mobasher}, \& {Siana}}]{topping20a}
---. 2020{\natexlab{b}}, \mnras, 495, 4430

\bibitem[{{Trebitsch} {et~al.}(2017){Trebitsch}, {Blaizot}, {Rosdahl},
  {Devriendt}, \& {Slyz}}]{trebitsch17}
{Trebitsch}, M., {Blaizot}, J., {Rosdahl}, J., {Devriendt}, J., \& {Slyz}, A.
  2017, \mnras, 470, 224

\bibitem[{{Troncoso} {et~al.}(2014){Troncoso}, {Maiolino}, {Sommariva},
  {Cresci}, {Mannucci}, {Marconi}, {Meneghetti}, {Grazian}, {Cimatti},
  {Fontana}, {Nagao}, \& {Pentericci}}]{troncoso14}
{Troncoso}, P., {Maiolino}, R., {Sommariva}, V., {et~al.} 2014, \aap, 563, A58

\bibitem[{{van der Wel} {et~al.}(2014){van der Wel}, {Chang}, {Bell}, {Holden},
  {Ferguson}, {Giavalisco}, {Rix}, {Skelton}, {Whitaker}, {Momcheva},
  {Brammer}, {Kassin}, {Martig}, {Dekel}, {Ceverino}, {Koo}, {Mozena}, {van
  Dokkum}, {Franx}, {Faber}, \& {Primack}}]{vanderwel14}
{van der Wel}, A., {Chang}, Y.-Y., {Bell}, E.~F., {et~al.} 2014, \apjl, 792, L6

\bibitem[{{Vink}(2022)}]{vink22}
{Vink}, J.~S. 2022, \araa, 60, 203

\bibitem[{{Vink} {et~al.}(2001){Vink}, {de Koter}, \& {Lamers}}]{vink01}
{Vink}, J.~S., {de Koter}, A., \& {Lamers}, H.~J.~G.~L.~M. 2001, \aap, 369, 574

\bibitem[{{Williams} {et~al.}(2022){Williams}, {Kelly}, {Chen}, {Brammer},
  {Zitrin}, {Treu}, {Scarlata}, {Koekemoer}, {Oguri}, {Lin}, {Diego}, {Nonino},
  {Hjorth}, {Langeroodi}, {Broadhurst}, {Rogers}, {Perez-Fournon}, {Foley},
  {Jha}, {Filippenko}, {Strolger}, {Pierel}, {Poidevin}, \&
  {Yang}}]{williams22}
{Williams}, H., {Kelly}, P.~L., {Chen}, W., {et~al.} 2022, arXiv e-prints,
  arXiv:2210.15699

\bibitem[{{Yeh} \& {Matzner}(2012)}]{yeh12}
{Yeh}, S. C.~C., \& {Matzner}, C.~D. 2012, \apj, 757, 108

\end{thebibliography}

\end{document}